\documentclass[sigconf]{acmart}

\AtBeginDocument{%
  }

\copyrightyear{2025}
\acmYear{2025}
\setcopyright{cc}
\setcctype{by}
\acmConference[MM '25]{Proceedings of the 33rd ACM International Conference on Multimedia}{October 27--31, 2025}{Dublin, Ireland}
\acmBooktitle{Proceedings of the 33rd ACM International Conference on Multimedia (MM '25), Oct. 27--31, 2025, Dublin, Ireland}\acmDOI{10.1145/3746027.3754752}
\acmISBN{979-8-4007-2035-2/2025/10}

\acmSubmissionID{MM-446}

\usepackage{multirow}
\usepackage{xcolor} 
\usepackage{colortbl} 
\usepackage[most]{tcolorbox}
\usepackage{fontawesome}
\newcommand{\model}{NEXUS-O}
\begin{document}
\title{\model: An Omni-Perceptive And -Interactive Model for Language, Audio, And Vision}

\author{$\text{Che Liu}^{*}$}
\orcid{0009-0004-3738-7998}
\affiliation{
 \institution{Imperial College London}
 \city{London}
 \country{UK}}
\email{che.liu21@imperial.ac.uk}

\author{$\text{Yingji Zhang}^{*}$}
\orcid{0000-0003-1499-3309}
\affiliation{
 \institution{University of Manchester}
 \city{Manchester}
 \country{UK}}
\email{yingji.zhang@manchester.ac.uk}

\author{\text{Dong Zhang*} \\ \text{Weijie Zhang*} \\ \text{Chenggong Gong*}}
\orcid{0009-0001-8811-8454}
\affiliation{
 \institution{HiThink Research}
 \city{Hangzhou}
 \country{China}}
 \email{}

\author{$\text{Yu Lu}$}
 \orcid{0009-0003-9695-6241}
\affiliation{
 \institution{HiThink Research}
 \city{Hangzhou}
 \country{China}}
 \email{luyu2@myhexin.com}

\author{Shilin Zhou}
\orcid{0009-0005-4500-3963}
\affiliation{
 \institution{Soochow University}
 \city{Suzhou}
 \country{China}}
 \email{slzhou.cs@outlook.com}

 \author{$\text{Ziliang Gan}$}
\orcid{0009-0000-7450-9887}
\affiliation{
 \institution{HiThink Research}
 \city{Hangzhou}
 \country{China}}
\email{ganziliang@myhexin.com}

\author{Ziao Wang}
\orcid{0000-0002-8019-2334}
\affiliation{
 \institution{Hong Kong Baptist University}
 \city{Hong Kong}
 \country{China}}
\email{wangziao1993@hotmail.com}

\author{\text{Haipang Wu} \\ \text{Ji Liu}}
\orcid{}
\affiliation{
 \institution{HiThink Research}
 \city{Hangzhou}
 \country{China}}
\email{}

\author{Andre Freitas}
\orcid{0000-0002-4430-4837}
\affiliation{
 \institution{University of Manchester}
 \city{Manchester}
 \country{UK}}
 \affiliation{
 \institution{Idiap Research Institute}
 \city{Marigny}
 \country{Swizerland}}
\email{andre.freitas@manchester.ac.uk}

\author{Qifan Wang}
\orcid{0000-0002-7570-5756}
\affiliation{
\institution{Meta AI}
\city{Menlo Park}
\country{USA}}
\email{wqfcr@fb.com}

\author{Zenglin Xu}
\orcid{0000-0001-5550-6461}
\affiliation{
\institution{Fudan University}
\city{Shanghai}
\country{China}}
\email{zenglin@gmail.com}

\author{Rongjunchen Zhang$\dagger$}
\orcid{0000-0003-1823-2726}
\affiliation{
\institution{HiThink Research}
\city{Hangzhou}
\country{China}}
\email{zhangrongjunchen@myhexin.com}

\author{$\text{Yong Dai}^{\dagger}\ddagger$}
\orcid{0000-0002-3041-5851}
\affiliation{
 \institution{X-Humanoid}
 \city{Beijing}
 \country{China}}
 \affiliation{
\institution{Fudan University}
 \city{Shanghai}
 \country{China}}
\email{daiyongya@outlook.com}
\renewcommand{\shortauthors}{Che Liu et al.}
\begin{abstract}

Human beings perceive the real world through a spectrum of sensory modalities, encompassing auditory, visual, and linguistic faculties. This work proposes an industry-level omni-modal large language model (LLM) pipeline that integrates auditory, visual, and linguistic modalities to overcome challenges such as limited tri-modal datasets, high computational costs, and complex feature alignments. Our pipeline consists of three main components: First, a modular, end-to-end framework enabling flexible configuration of various encoder–LLM–decoder architectures. Second, a lightweight training strategy that pre-trains audio-language alignment on the state-of-the-art vision-language model Qwen2.5-VL, thus avoiding the costly pre-training of vision-specific modalities. Third, an audio synthesis pipeline that generates high-quality audio-text data from diverse real-world scenarios, supporting applications such as Automatic Speech Recognition and Speech-to-Speech chat.
To this end, we introduce an industry-level omni-modal LLM, \textbf{\model}. Extensive experiments validate the efficacy of our pipeline, yielding the following key findings: 
(1) In the visual understanding task, \model~exhibits superior performance compared with its backbone model - Qwen2.5-VL-7B, validating the efficiency of our training strategy. (2) Within the English Spoken Question-Answering task, the model achieves better accuracy than the same-period competitor (i.e, MiniCPM-o2.6-7B) in the LLaMA Q. benchmark. 
(3) In our real-world ASR testset, \model{} achieves outstanding performance, indicating its robustness in real scenarios. (4) In the Speech-to-Text Translation task, our model outperforms Qwen2-Audio-Instruct-7B. (5) In the Text-to-Speech task, based on pretrained vocoder (e.g., Fishspeech1.4 or CosyVoice2.0), \model{} is comparable to its backbone vocoder on Seed-TTS benchmark. (6) An in-depth analysis of tri-modal alignment reveals that incorporating the audio modality enhances representational alignment between vision and language.
\end{abstract}

\ccsdesc[500]{Computing methodologies~Natural language generation}
\keywords{Omni-modalities, Mutimodal Large Language Models}

\maketitle

\def\customfootnotetext#1#2{{%
  \let\thefootnote\relax
  \footnotetext[#1]{#2}}}
\customfootnotetext{1}{$\dagger$ denotes project leader.}
\customfootnotetext{1}{\textsuperscript{*} denotes equal contribution.}
\customfootnotetext{1}{$\ddagger$ denotes corresponding author.}

\section{Introduction}
Since the mid-20th century, significant efforts have been devoted to designing systems of the capability to simulate human cognition, with the ultimate goal of developing Artificial General Intelligence (AGI). Notable algorithms include Rule-based systems, Neuro-symbolic AI systems, Neural networks systems, etc. Among them, neural networks systems have emerged as a cornerstone thanks to their remarkable generalization capabilities. In this context, various neural architectures have been proposed, such as CNN and Transformer. The discovery of neural scaling laws~\citep{kaplan2020scalinglawsneurallanguage} has further advanced the field, paving the way for Transformer-based Large Language Models (LLMs), such as ChatGPT, which centre on language as the medium for perceiving and interacting with both human society and the physical world. By aligning multiple modalities around the language, Multimodal Large Language Models (MLLMs) have gained widespread recognition as a promising pathway toward AGI.

MLLMs, involving language, vision, and auditory modalities, can generally be divided into three categories: vision-language model \citep{wang2024qwen2}, audio-language model \citep{chu2024qwen2}, and vision-language-audio model \citep{fu2025vita}. While alignment among three modalities offers a more comprehensive simulation of human cognitive processes than two-modality alignment, it presents significantly greater challenges. These challenges include limited accessibility to tri-modal datasets, higher computational resource requirements, and the complexity of aligning features across three modalities. These limitations hinder the advancement of omni-perceptive and interactive AGI systems. To address these challenges, we propose \textit{a comprehensive, industry-level omni-modal LLM pipeline that comprises a modularised end-to-end framework, a lightweight training strategy, and an audio data synthesis pipeline}:

First, in the modularised framework, each modality (audio, image/video, and text) is processed using its own dedicated pretrained encoder. This design ensures that the inherent characteristics and fine-grained details unique to each modality are effectively preserved within their respective latent representations. Subsequently, these modality-specific representations are integrated and aligned within a LLM, and the resulting unified representations are subsequently supplied to decoupled decoders designed for each modality. This modular formalism design facilitates efficient and flexible configuration of various encoder–LLM–decoder architectures, thereby potentially contributing to the advancement of AGI by integrating components such as language models, world models, and embodied intelligence.

Second, most omni-modal models, such as Qwen2.5-omni\cite{xu2025qwen25omnitechnicalreport}, VITA1.5\cite{fu2024vita}, etc., necessitate a preliminary pre-training stage for vision alignment, understanding, and reasoning. This stage typically demands substantial computational resources due to the processing of long sequences of image tokens, which is consequently less accessible to smaller research institutions and companies. To mitigate this limitation, we only pre-train audio-language alignment over the current state-of-the-art vision-language model, Qwen2.5-VL~\citep{bai2025qwen25vltechnicalreport}\footnote{In this report, Qwen2.5-VL refers to the instruction version.}. Due to the inherent alignment between audio and text modalities, this training strategy effectively integrates audio understanding and generation capabilities while preserving the model's established vision comprehension and reasoning functions.

Third, high-quality audio training data are often less accessible in both academy and industry. To address this limitation, we propose an audio synthesis pipeline that generates high-quality audio-text data spanning a broad spectrum of real-world scenarios, such as live streaming and dialogue, thereby supporting various downstream tasks such as Automatic Speech Recognition (ASR) and Speech-to-Speech chat. By incorporating real-world audio data at the pretraining stage, the model can be fast deployed for real-world applications. Based on our audio synthesis pipeline, we introduce a real-life ASR testset, named as \model{}-audio.

To this end, we introduce \textbf{\model{}}\footnote{“Nexus”: connect the strongest models to form an even more powerful integrated model.}, an industry-level omni-modal LLM. Extensive experiments validate the efficacy of our pipeline, yielding the following key findings: 
(1) In the visual understanding task, \model~exhibits superior performance compared with its backbone model - Qwen2.5-VL-7B, validating the efficiency of our training strategy. (2) Within the English Spoken Question-Answering task, the model achieves better accuracy than the same-period competitor (i.e, MiniCPM-o2.6-7B) in the LLaMA Q. benchmark \citep{nachmani2024spokenquestionansweringspeech}. 
(3) In our real-world ASR testset, \model{} achieves outstanding performance, indicating its robustness in real scenarios. (4) In the Speech-to-Text Translation task, our model outperforms Qwen2-Audio-Instruct-7B. (5) In the Text-to-Speech task, based on pretrained vocoder (e.g., Fishspeech1.4 or CosyVoice2.0) \model{} is comparable to its backbone vocoder on Seed-TTS benchmark \cite{anastassiou2024seedttsfamilyhighqualityversatile}. 

To further investigate the interplay among three modalities, we conducted an in-depth analysis of tri-modal alignment. Our findings indicate that (6) the inherent correspondence between audio and language can serve to facilitate and reinforce vision-language alignment, indicating the efficiency of our training strategy. The main contributions of this paper are summarised as follows:

\begin{itemize}
\item We propose a comprehensive, industry-level omni-modal LLM pipeline that comprises a modularised end-to-end framework, a lightweight training strategy, and an audio data synthesis pipeline.

\item Based on the proposed pipeline, we introduce \model, an industry-level omni-modal LLM. It is designed to support any combination of audio, image/video, and text inputs, and it is capable of generating outputs in either audio or language modalities. The pretrained model is available at \faGithub \textit{ } https://github.com/HiThink-Research/NEXUS-O

\item We systematically design the speech data synthesis pipeline to obtain high-quality, real-life speech datasets, covering various real-world scenarios, such as corporate meetings, live broadcasting scenarios, etc., to facilitate the rapid deployment of the model in real-world scenarios. Moreover, we introduce the \model-audio testset to evaluate its robustness in real life ASR task.

\item We comprehensively evaluate the performance of \model~on different benchmarks and conduct an in-depth analysis of vision-language alignment by incorporating audio modality within the representation space. Experimental results indicate the merits of our proposed pipeline.
\end{itemize}



\section{Related Work} \label{sec:relate}

In this section, we present the existing works of Multimodal Large Language Models (MLLMs) in Section \ref{sec:relate_mllms}, and analyse existing audio modality evaluation approaches in Section \ref{sec:relate_evaluation}.

\subsection{MLLMs} \label{sec:relate_mllms}

Large Language Models (LLMs) have demonstrated remarkable capabilities in understanding and reasoning over textual knowledge \citep{yang2024qwen2}. Building on these advancements, a bunch of studies \citep{dai2022one} have extended the understanding and alignment capabilities of LLMs to visual knowledge, addressing challenges such as vision-language alignment and instruction-following, by introducing Multimodal Large Language Models (MLLMs). Notable models include Qwen-VL \citep{wang2024qwen2}, InternVL \citep{chen2024internvl}, etc. However, in complex visual reasoning tasks, such as video analysis, MLLMs typically struggle to perform well due to the heterogeneity among diverse modalities, where different modalities have distinct feature granularities, representations, and geometries in a shared latent space \citep{zhang-etal-2024-graph}.
To further improve their performance in visual understanding tasks, recent works have expanded MLLMs by incorporating audio capabilities, such as LLaMA-Omni \citep{fang2024llama}, VITA-1.0 \citep{fu2024vita}, Mini-Omni2 \citep{xie2024mini2}, Baichuan-Omni \citep{li2024baichuan}, and MinMo \citep{chen2025minmomultimodallargelanguage}. However, these models exhibit certain limitations: LLaMA-Omni and MinMo lacks visual capabilities, Baichuan-Omni does not have end-to-end Text To Speech (TTS) capabilities, and VITA-1.0 is constrained by the limited capability of its backbone model.

Recent models, such as VITA-1.5~\citep{fu2025vita}, MiniCPM-o2.6~\footnote{https://github.com/OpenBMB/MiniCPM-o}, Baichuan-omni-1.5\footnote{\url{https://github.com/baichuan-inc/Baichuan-Omni-1.5}}, and OpenOmni \citep{luo2025openomnilargelanguagemodels}, have been proposed and have achieved state-of-the-art performance. However, these models usually require a visual pretraining stage. While pretraining from scratch can enhance multimodal alignment, this strategy significantly increases training time and resource demands due to long sequence of image or video tokens and large pretraining vision data, which is less friendly with institutions with limited computation resources. To improve training efficiency, we start by pre-training over MLLM, specifically Qwen2.5-VL-7B. Owing to the natural alignment between audio and language modalities, the integration of the audio modality has minimal impact on the pretrained geometric and manifold structures underlying the vision-language alignment. As a result, the model’s established capabilities in vision comprehension and reasoning are effectively preserved. 

\subsection{Audio Modality Evaluation} \label{sec:relate_evaluation}

With the rapid advancement of MLLMs, a wide range of multimodal benchmarks has been proposed and widely adopted for evaluation, ranging over vision \citep{fu2024videommefirstevercomprehensiveevaluation}, audio \citep{fu2024mmecomprehensiveevaluationbenchmark}, omni \citep{li2024omnibenchfutureuniversalomnilanguage}, etc. In the speech domain, existing benchmarks, e.g., Fleurs \citep{conneau2023fleurs}, Aishell2 \citep{du2018aishell}, LibriSpeech \citep{pratap2020mls}, and Common Voice \citep{ardila2019common}, provide linguistically diverse datasets under varying acoustic conditions, facilitating Automatic Speech Recognition (ASR) evaluation. Additionally, AIR-Bench \citep{yang2024air} introduces a pioneering generative evaluation framework encompassing speech, natural sounds, and music, employing novel audio mixing strategies and GPT-4 for unified, objective, and reproducible assessments.
However, the assessment of speech models should be closely aligned with real-world environments to ensure that the evaluation results faithfully represent the demands and requirements of practical and industrial applications. We observe that these existing speech benchmarks present three limitations. First, the small data scale of these benchmarks typically results in significant performance variance. Second, the benchmarks primarily focus on controlled scenarios, making it difficult to evaluate models in dynamic and complex real-world settings, e.g., meetings or live broadcast environments. Third, quantitative evaluation scores from the existing ASR benchmarks generally fail to align closely with the actual user interaction experience, thereby limiting their practical utility. To address these limitations, we propose \model-audio testset, {\color{black}comprising 8.1k and 2.8k of Chinese and English ASR samples}, spanning various application domains, such as corporate meetings and live broadcasting scenarios.



\begin{figure}[t]
    \centering
    \includegraphics[width=\columnwidth]{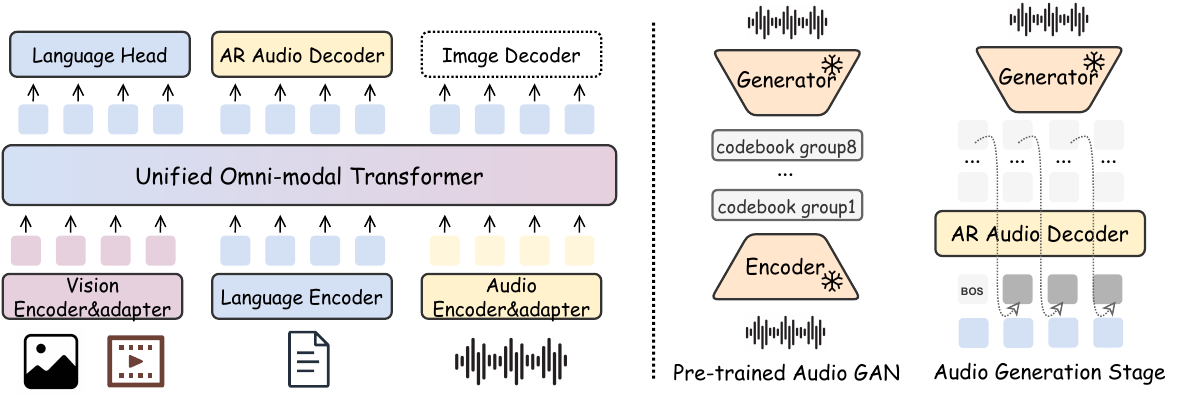}
    \caption{Modularised Architecture, which is designed to accept any combination of input modalities and generates output in either the language or audio modality, where the Auto-Regressive (AR) audio decoder takes a special start token embedding and the last language embedding as input to generate the hierarchical discrete audio codes in an auto-regressive manner. These codes are subsequently fed into a pretrained audio generator to produce the final audio output.}
    \label{fig:architecture}
\end{figure}

\section{\model} \label{sec:model}
In this section, we introduce the architecture of \model~in Section~\ref{sec:arc}, training data composition and audio data synthesis pipeline in Section  \ref{sec:data}, pre-training strategies in Section \ref{sec:tr}, respectively.
\subsection{Modularised Architecture} \label{sec:arc}

\paragraph{Visual encoder.} 
Following Qwen2.5-VL \citep{bai2025qwen25vltechnicalreport}, the vision encoder employs a re-designed Vision Transformer (ViT) architecture by incorporating M-RoPE and window attention to support native input resolutions while accelerating the computation of the entire visual encoder. During both training and inference, the height and width of the input images are resized to multiples of 28 before being fed into the ViT. The vision encoder processes images by splitting them into patches with a stride of 14, generating a set of image features.

\paragraph{Audio encoder/decoder.} To enable audio capabilities in the vision-language MLLM, we incorporate an audio encoder-decoder architecture. In this setup, the encoder is responsible for mapping the speech features into the semantic space of the MLLM, while the decoder transforms the semantic code back into speech, as shown in Figure \ref{fig:architecture}. Specifically, the audio encoder comprises a pre-trained Whisper-large-v3 \citep{radford2023robust}, and a two-layer MLP adapter. Specifically, the first layer of the MLP concatenates five consecutive speech feature vectors into a single vector, which is subsequently processed by the second MLP layer.

The auto-regressive audio decoder is a 6-layer decoder-only transformer designed to generate discrete speech codes, which are then passed to a pre-trained generator to produce the final waveforms. Specifically, we use the audio codebook from the FireFly GAN in Fish-Speech \citep{liao2024fishspeechleveraginglargelanguage} as the audio codebook. Following MusicGen \citep{copet2023simple}, the audio decoder predicts sequences of eight speech codes in a delayed manner. At each time step, the decoder averages the embeddings of the eight speech codes generated in the previous time step with the hidden state of the LLM at the current time step. The resulting averaged embedding is then fed into the audio transformer to predict the next eight speech tokens. Finally, the predicted speech tokens are input into the generator of the FireFly GAN to synthesise the final waveforms.

\paragraph{Language model.} \model~utilises the language model component of Qwen2.5-VL-7B \citep{bai2025qwen25vltechnicalreport} as the foundational language model.
\subsection{Training Data} \label{sec:data}
\begin{figure}[t]
    \centering
    \includegraphics[width=\columnwidth]{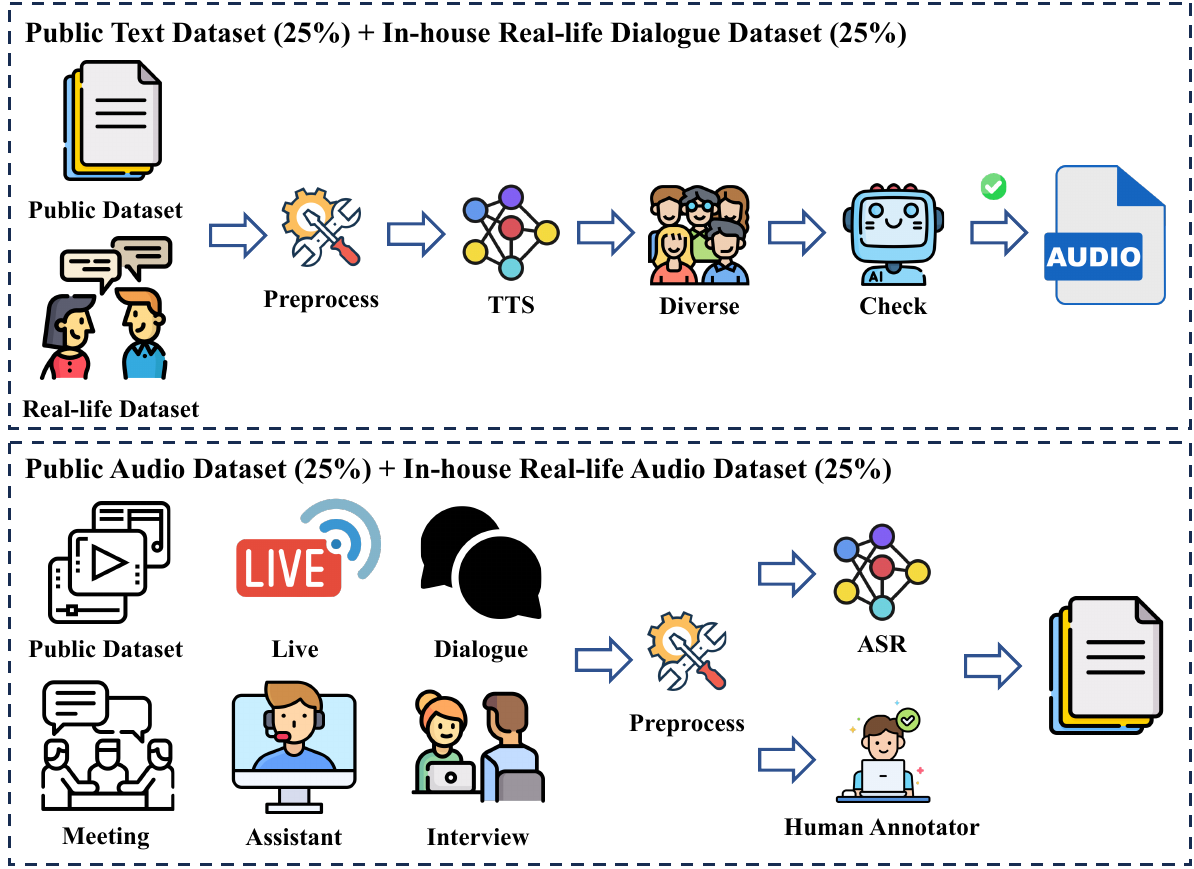}
    \caption{Audio dataset synthesis pipeline, the upper component represents the text-to-audio branch, while the lower component corresponds to the audio-to-text branch. Both components incorporate an equal proportion of in house, real-world samples.}
    \label{fig:data}
\end{figure}
\paragraph{Datasets composition.} As detailed in Table~\ref{tab:dataset}, the training data for \model{} encompasses a wide range of domains within speech processing. These domains include (i) recognition tasks such as ASR and Speech Emotion Recognition (SER), (ii) reasoning tasks include Automatic Audio Caption and Speech Emotion Recognition, (iii) translation tasks like Speech-to-Text Translation (S2TT), generation tasks such as Speech Synthesis, and (iv) interactive tasks involving Speech-to-Speech Chat and Speech-to-Text Chat. The pretrained datasets include (1) public corpora (50\%): Aishell1 \citep{bu2017aishell1opensourcemandarinspeech}, Wenetspeech \citep{zhang2022wenetspeech}, Librispeech \citep{7178964}, Kensho \citep{oneill2021spgispeech5000hourstranscribed}, PeopleSpeech \citep{galvez2021peoplesspeechlargescalediverse}, MLS \citep{Pratap2020MLSAL}, and Gigast \citep{,ye2023gigast10000hourpseudospeech} and (2) real-life in-house data (50\%).

Table \ref{tab:dataset} presents the statistical information for the audio training dataset. Unlike most Omni-MLLMs, which typically report duration in hours \cite{chen2025minmomultimodallargelanguage}, our summary is based on the number of audio samples. The total duration of the dataset is approximately 30k hours, this is considerably smaller than the durations employed in other works, such as MinMo (pretrained over 1.4 million hours) \cite{chen2025minmomultimodallargelanguage} and Qwen2-Audio (pretrained over 520k hours) \cite{chu2024qwen2audiotechnicalreport}.
\begin{table}[h!]
\caption{{\color{black}Datasets description.}
}

\label{tab:dataset}
\centering
\resizebox{\columnwidth}{!}{
\begin{tabular}{clcc}
\toprule
\textbf{Stage} & \textbf{Task Description}  & \textbf{Num} & \textbf{GPUs (H100)} \\ \midrule

\multirow{5}{*}{1}
& Automatic Speech Recognition & 5.6M  & \multirow{5}{*}{6} \\
& Speech-to-Text Translation  & 1.5M  \\
& Speech-to-Text Chat & 0.2M  \\
& Automatic Audio Caption & 0.14M   \\
& Speech Emotion Recognition & 0.75M   \\
\hline
\multirow{3}{*}{2}
& Speech-to-Text Chat & 0.23M  & \multirow{3}{*}{64} \\
& Automatic Speech Recognition & 20k  \\
& Speech-to-Text Translation  & 20k  \\
\hline 
\multirow{2}{*}{3}
& Speech Synthesis & 40M  & \multirow{2}{*}{6} \\
& Speech-to-Speech chat & 0.22M  & \\

\bottomrule
\end{tabular}
}
\end{table}

\paragraph{Audio dataset synthesis pipeline.} We integrate both English and Chinese speech and textual datasets to generate paired speech-text data for ASR, speech synthesis (text-to-speech), Speech-to-Text Chat, and Speech-to-Speech Chat tasks. For the speech corpus (depicted at the bottom of Figure \ref{fig:data}), we employ Whisper \citep{radford2022robustspeechrecognitionlargescale} to transcribe speech into text. For noisy samples in our in-house dataset, human annotators are engaged to ensure accurate transcription.

For the textural corpus (top in Figure \ref{fig:data}), we employ the zero-shot in-context generation technique proposed by CosyVoice \citep{du2024cosyvoice} to convert the text into natural-sounding speech. However, many conversational and instruction textural datasets are not ideal for speech synthesis due to the nature of the tasks, which do not facilitate effective speech interaction. To address this issue, we enhance the quality of our synthesised speech data by implementing the following pre-processing rules:
\begin{enumerate}
\item Length Filtering: texts exceeding a length threshold are excluded from the dataset. Specifically, texts longer than 200 Chinese characters or 200 English words are removed.
\item Non-text Element Filtering: texts containing an excessively high proportion of non-text elements are excluded. Specifically, texts with a non-text character ratio exceeding 0.2 are removed. Text characters in this context include punctuation marks (e.g., !), alphabetic characters ([a-z]), and Chinese characters.
\item Pattern Matching Filtering: texts that included URLs, file paths, or common LaTeX formulas were identified and removed using regular expressions. 
\end{enumerate}
For Speech-to-Text Chat and Speech-to-Speech Chat tasks, some texts are unsuitable for a conversational context. To address this limitation, we implement the following rules:
\begin{enumerate}
\item Excluding queries that are not appropriate for the Chat task. For example, questions involving metathetical questions are filtered out.
\item Rewriting the question in a conversational tone and generating a conversational response via Qwen2-72B \citep{chu2024qwen2}.
\end{enumerate}
To enhance the diversity of voices, we randomly select a speaker from a pool of two thousand individuals for each corresponding language to synthesise the speech, ensuring that the voice aligns with the language of the text. Following the text-to-speech (TTS) process, an automated verification procedure is executed using the pretrained Whisper model, which checks whether the synthesised sample can be approximately reconverted to the original source sample, thereby ensuring the quality of the data synthesis.

Furthermore, we introduce the \textbf{Nexus-audio testset}, which comprises 8.1k Chinese and 2.8k English ASR samples. This testset spans a range of application domains, such as corporate meetings, live broadcasting scenarios, etc.

\subsection{Training Strategies} \label{sec:tr}
\paragraph{Stage 1: audio alignment.} The first stage aims to facilitate the capacity of LLMs to comprehend and process input speech features. {\color{black}To accomplish this, we utilise 8 million bilingual speech-text pairs for alignment training, covering five training objectives, as illustrated in Table \ref{tab:dataset}.} 






\paragraph{Stage 2: audio SFT} The second stage aims to enable the model to follow speech instructions by incorporating speech-based questions and text-based answers. In this stage, we freeze the visual and audio encoders while unfreezing the visual and audio adapters, and backbone LLM. This strategy is designed to enhance the capability of \model{} to respond accurately to multimodal instructions, thereby improving its performance in understanding and processing both visual and auditory information.

By keeping the core encoder parameters fixed, we ensure that the foundational representations remain stable while allowing the adapters to fine-tune and specialize in the nuances of specific tasks or instructions. This strategic adjustment aims to refine the sensitivity of \model{} to multimodal cues without disrupting the robustness of its underlying architecture.

\paragraph{Stage 3: audio output.}
Different from the previous large speech model connected to the TTS module \citep{fu2024vita}, we exploit a decoder to generate speech end-to-end. First, we train \model{} on the Text-to-Speech dataset to make the model learn text-to-speech alignment. Then, we utilize the speech-to-speech dataset to train the model to learn speech understanding and speech synthesis capabilities. At this stage, only the parameters of the audio decoder are trainable.

\section{Evaluation} \label{sec:eval}
We systematically evaluate our model performance from two perspectives: (i) downstream tasks alignment analysis in Section~\ref{sec:eval_task_align} and (ii) representation space alignment analysis in Section~\ref{sec:eval_embed_align}. 

\subsection{Multimodal Tasks Alignment Analysis} \label{sec:eval_task_align}

In this section, we evaluate the performance of \model{} on the following five tasks: vision understanding, Spoken English QA tasks, ASR, Speech-to-Text translation, and Text-to-Speech tasks. These tasks cover a wide range of modalities and capabilities, providing a comprehensive assessment of the versatility and effectiveness of MLLMs across various domains.

\paragraph{\textbf{Vision-Language Evaluation}} \label{sec:vision_language}
We conduct a comprehensive quantitative evaluation of \model on a range of vision understanding tasks using eight distinct benchmarks:
(1) HallusionBench (Hal) \citep{guan2023hallusionbench}: Evaluates the model's susceptibility to hallucination and its handling of visual illusions.
(2) MathVista (MathV) \citep{lu2023mathvista}: Assesses mathematical reasoning within visual contexts.
(3) OCRBench (OCR) \citep{fu2024ocrbenchv2improvedbenchmark}: Measures optical character recognition (OCR) capabilities on text-intensive images.
(4) Video-MME \citep{fu2024videommefirstevercomprehensiveevaluation}: Examines video question-answering reasoning abilities across six domains over diverse temporal ranges.
(5) MMMU \citep{yue2024mmmumassivemultidisciplinemultimodal}: Challenges the model with multi-disciplinary tasks that demand college-level subject knowledge and deliberate reasoning.
(6) AI2D: A dataset comprising over 5,000 grade school science diagrams.
(7) MMVet \citep{yu2024mmvetevaluatinglargemultimodal}: Evaluates integrated capabilities including recognition, OCR, domain knowledge, language generation, spatial awareness, and mathematical reasoning.
(8) MME \citep{fu2024mmecomprehensiveevaluationbenchmark}: Measures both perceptual and cognitive abilities across 14 subtasks.
\begin{table*}[h!]
\caption{\textbf{Evaluation on Vision Understanding Benchmarks.} The best two values are shown in \textbf{\textcolor{black}{bold}} and \underline{underlined}. The blue row refers to the main competitors. In MMMU benchmark, * indicates that the measurement is developed over the validation set. We can observe that \model~shows performance comparable to the leading open-source models and advanced closed-source counterparts.}
    \label{fig:image}
\resizebox{\textwidth}{!}{
\begin{tabular}{llcccccccc}
\toprule
\textbf{Model} & \textbf{LLM-size}   & \textbf{Video-MME}  & \textbf{MMMU} & \textbf{MathV}   & \textbf{Hal} & \textbf{AI2D} & \textbf{OCR} & \textbf{MMVet} & \textbf{MME}\\ \hline
\multicolumn{10}{c}{{\cellcolor[rgb]{0.957,0.957,0.957}} \textbf{\textit{Vision-Language Models}}} \\
VILA-1.5\cite{lin2024vilapretrainingvisuallanguage} & Vicuna-v1.5-13B &  44.2 & 41.1 & 42.5 & 39.3 & 69.9 & 460.0 & 45.0 & 1718.2\\
LLaVA-Next\cite{liu2024llavanext} & Yi-34B  & 51.6 & 48.8 & 40.4 & 34.8 & 78.9 & 574.0 & 50.7 & 2006.5 \\
InternLM-Xcomposer2\cite{dong2024internlmxcomposer2masteringfreeformtextimage} & InternLM2-7B  & 56.2 & 41.4 & 59.5 & 41.0 & 81.2 & 532.0 & 46.7 & 2220.4 \\
Cambrian\cite{tong2024cambrian1fullyopenvisioncentric} & NousHermes2-Yi-34B  & 54.2 & 50.4 & 50.3 & 41.6 & 79.5 & 591.0 & 53.2 & 2049.9 \\
InternVL-Chat-1.5\cite{chen2024fargpt4vclosinggap} & InternLM2-20B  & 57.1 & 46.8 & 54.7 & 47.4 & 80.6 & 720.0 & 55.4 & 2189.6 \\
Ovis1.5\cite{lu2024ovisstructuralembeddingalignment} & Gemma2-It-9B  & 58.1 & 49.7 & 65.6 & 48.2 & \textbf{84.5} & 752.0 & 53.8 & 2125.2 \\
InternVL2\cite{chen2024internvl} & InternLM2.5-7B  & \underline{61.5} & 51.2 & 58.3 & 45.0 & \underline{83.6} & 794.0 & 54.3 & 2215.1 \\
\rowcolor{blue!8}{MiniCPM-V 2.6\cite{yao2024minicpm}} & Qwen2-7B  & 57.5 & 49.8 & 60.6 & 48.1 & 82.1 & 852.0 & 60.0 & 2268.7 \\
\rowcolor{blue!8}Qwen2.5-VL\cite{bai2025qwen25vltechnicalreport} & Qwen2.5-7B  & 56.0 & 51.8* & 61.1 & \textbf{71.7} & 80.7 & \underline{877.0} & - & 2299.1 \\
\hline
\multicolumn{10}{c}{{\cellcolor[rgb]{0.957,0.957,0.957}} \textbf{\textit{Omni-modal Models}}} \\
\rowcolor{blue!8}{VITA-1.5-Audio\cite{fu2025vita15gpt4olevelrealtime}} & Qwen2-7B  & - & 52.1 & \textbf{66.2} & 44.9 & 79.3 & 732.0 & 49.6 & \textbf{2352.0} \\
EMova-8B\cite{chen2025emovaempoweringlanguagemodels} & LLaMA-3.1-8B & - & - & 61.1 & -& 82.8 & 824.0 & 55.8 & 2205.0 \\
Baichuan-Omni-1.5\cite{li2024baichuan} & - & 58.2 & 47.3 & 51.9 & 47.8 & - & - & 65.4 & 2186.9 \\
Megrez-3B-Omni\cite{li2025megrezomnitechnicalreport} & Megrez-3B  & - & 51.8 & 62.0 & 50.1 & 82.0 & - & - & 2315.0 \\
\hline
\multicolumn{10}{c}{{\cellcolor[rgb]{0.957,0.957,0.957}} \textbf{\textit{Proprietary}}} \\
GPT-4V & -  & 50.4 & 59.3 & 48.2 & 39.3 & 71.4 & 678.0 & 49.0 & 1790.3 \\
GPT-4o mini  & - & 54.8 & 60.0 & 52.4 & 46.1 & 77.8 & 785.0 & \textbf{66.9} & 2003.4 \\
Gemini 1.5 Pro & 200B  & 59.1 & 60.6 & 57.7 & 45.6 & 79.1 & 754.0 & 64.0 & 2110.6\\
GPT-4o  & - & 61.6 & \underline{62.8} & 56.5 & 51.7 & 77.4 & 663.0 & \underline{66.5} & \underline{2328.7}\\
Claude3.5 Sonnet & 175B  & \textbf{62.2} & \textbf{65.9} & 61.6 & 49.9 & 80.2 & 788.0 & 66.0 & 1920.0 \\ \hline \rowcolor[gray]{0.9}
\multicolumn{10}{c}{{\cellcolor[rgb]{0.957,0.957,0.957}} \textbf{\textit{Our Model}}} \\
\rowcolor{blue!8}{\model} & Qwen2.5-VL-7B  & 57.0 & 53.2* & \underline{62.1} & \underline{71.1} & 81.2 & \textbf{882.0} & - & 2315.5 \\ \bottomrule
\end{tabular}
}
\end{table*}

As illustrated in Table \ref{fig:image}, we observe that \model\ achieves strong performance to current popular vision-language MLLMs on the MathV, OCR, and Hal benchmarks. Furthermore, when compared to Qwen2.5-VL, \model\ demonstrates superior performance, highlighting its competitive capabilities in vision understanding and reasoning tasks, indicating our model and training strategy can effectively maintain vision-language alignment capability by incorporating audio modality \textbf{(Finding1)}.

\paragraph{\textbf{Audio-Language Evaluation}} \label{sec:audio_language}

Next, we evaluate the audio-language alignment on Spoken Question-Answering (SQA), Automatic Speech Recognition (ASR), Speech-to-Text translation (S2TT), and Text-to-Speech (TTS) tasks. 

\paragraph{SQA task} We evaluate the performance on the English spoken QA benchmark: LLaMA Q. \citep{nachmani2024spokenquestionansweringspeech}. As presented in Table \ref{tab:eqa}, our model achieves top performance, higher than same-period competitor MiniCPM-o2.6-7B (highlighted in blue), demonstrating its competitive capabilities in this task. 
\begin{table}[ht!]
\caption{\textbf{Evaluation on Audio English QA Benchmarks.} The accuracy~(\%) of different models in English question answering on three sets. The parameter size is derived from the backbone LLMs. Our model achieves top accuracy in the LLaMA Q. benchmark, outperforming the same-period competitor MiniCPM-o2.6-7B.}
\label{tab:eqa}
\centering
\begin{tabular}{lcc}
\toprule
\textbf{Model}  & \textbf{Modality}  & \textbf{LLaMA Q.}$\uparrow$ \\ \hline
SpeechGPT-7B\citep{zhang2023speechgptempoweringlargelanguage} & Audio\&Text & 21.60 \\
Spectron-1B\citep{nachmani2023spoken} & Audio\&Text & 22.90                         \\
Moshi-7B\citep{defossez2024moshi}     & Audio\&Text & 62.30                         \\
GLM-4-Voice-9B\citep{zeng2024scalingspeechtextpretrainingsynthetic}  & Audio\&Text  & \underline{64.70}  \\
\rowcolor{blue!8}MiniCPM-o2.6-7B  & Audio\&Text & 61.00   \\
Mini-Omni-0.5B\citep{xie2024mini} & Audio\&Text & 22.00   \\
Llama-Omni-8B\citep{fang2024llamaomniseamlessspeechinteraction} & Audio\&Text  & 45.30 \\
\hline
\multicolumn{3}{c}{{\cellcolor[rgb]{0.957,0.957,0.957}} \textbf{\textit{Our Models}}} \\
\rowcolor{blue!8}\model  & Audio\&Text  & \textbf{67.33}  \\
\bottomrule
\end{tabular}
\end{table}

\paragraph{ASR task} In the ASR task, we focus on both Mandarin (CH) and English (EN), evaluating performance on the following benchmarks: AIShell-2 \citep{du2018aishell2transformingmandarinasr}, Librispeech \citep{panayotov2015librispeech}, and our real-scenario benchmark. As demonstrated in Table \ref{tab:asr}, our model achieves the best performance on the real scenario ASR testset, \model{}-audio, indicating the robustness of our model in real life \textbf{(Finding2)}.
\begin{table*}[h!]
\caption{\textbf{Evaluation on ASR Benchmarks, where CER and WER refer to Character Error Rate and Word Error Rate, respectively.} \model~has demonstrated strong performance in both real-life Mandarin and English ASR tasks. It outperforms specialized speech models, achieving better results in both languages. AIShell-2 is measured over all categories (Mac/iOS/Android), TC: test-clean, TO: test-other.} \label{tab:asr}
\begin{tabular}{lccccc} 
\toprule
\multirow{2}{*}{\textbf{Model}} & \multicolumn{2}{c}{\textbf{CH (CER$\downarrow$)}} & \multicolumn{3}{c}{\textbf{EN (WER$\downarrow$)}} \\ \cmidrule{2-6}
 & \textbf{AIShell-2} &  \textcolor{black}{\textcolor{black}{\textbf{\model-audio}}}   & \textbf{Librispeech TC} & \textbf{Librispeech TO}& \textcolor{black}{\textcolor{black}{\textbf{\model-audio}}} \\ \hline
\multicolumn{6}{c}{{\cellcolor[rgb]{0.957,0.957,0.957}} \textbf{\textit{Speech LLMs}}} \\
Qwen2-Audio-Instruct-7B\cite{chu2024qwen2audiotechnicalreport} & 3.00/3.00/2.90 &  35.45  & \textbf{1.70} & 4.00 & 26.12 \\ 
\multicolumn{6}{c}{{\cellcolor[rgb]{0.957,0.957,0.957}} \textbf{\textit{Omni-modal LLMs}}} \\
Mini-Omini2-0.5B\citep{xie2024mini}  &  &  117.71 & 9.80 & 4.70 & 54.69 \\
\rowcolor{blue!8}VITA-1.5-7B\citep{fu2025vita} & -  & 127.34  & 7.20 & \textbf{3.40} & 122.71 \\
MinMo-7B\citep{chen2025minmo} & 4.89/4.76/4.96  & -  &  \underline{1.74} &  \underline{3.89} & - \\
\rowcolor{blue!8}MiniCPM-o2.6-7B & -  & \underline{12.00}  &  \textbf{1.70} &  \underline{3.89} & \underline{18.00} \\
\hline
\multicolumn{6}{c}{{\cellcolor[rgb]{0.957,0.957,0.957}} \textbf{\textit{Our Models}}} \\

\rowcolor{blue!8}\textcolor{black}{\model}  & 3.35  & \textbf{11.82} & 3.50 & 5.29 & \textbf{12.32}  \\
\bottomrule
\end{tabular}
\end{table*}

\paragraph{S2TT task} As for the S2TT, we exploit CoVoST2 \citep{wang2020covost2massivelymultilingual}, which provides massive multilingual S2TT datasets. As shown in Table \ref{tab:speech2text}, \model~demonstrates superior performance compared to specialised speech LLM, Qwen2-Audio-7B-Instruct, in both translation tasks. Nevertheless, our model still falls short of the performance of its contemporary competitor, MiniCPM-o2.6-7B. However, we pre-train our model on a small audio dataset (about 30k hours) at the pre-training stage \textbf{(Finding3)}.
\begin{table}[h!]
\centering
\caption{\textbf{Evaluation on Speech-to-Text Translation Benchmarks.} 
\model~demonstrated superior performance compared to Qwen2-Audio-Instruct-7B in both translation tasks, however, it still falls short of the performance of its contemporary competitor due to the relatively modest scale of our pre-training speech dataset (about 30k hours).}\label{tab:speech2text}
\resizebox{8.5cm}{!}{
\begin{tabular}{lcc}
\toprule
\multirow{2}{*}{\textbf{Model}} & \multicolumn{1}{c}{\textbf{CH-EN (BLEU$\uparrow$)}} & \multicolumn{1}{c}{\textbf{EN-CH (BLEU$\uparrow$)}} \\ \cmidrule{2-3}
 & \textbf{CoVoST2}  & \textbf{CoVoST2} \\ \hline
 \multicolumn{3}{c}{{\cellcolor[rgb]{0.957,0.957,0.957}} \textbf{\textit{Speech LLMs}}} \\
Qwen2-Audio-7B\cite{chu2024qwen2audiotechnicalreport}  & 24.40 & 45.20 \\
Qwen2-Audio-Instruct-7B\cite{chu2024qwen2audiotechnicalreport}\textsuperscript{*} & 22.90 & 39.50 \\ 
MinMo-7B\cite{chen2025minmomultimodallargelanguage}\textsuperscript{*} & \underline{25.95} & \underline{46.68} \\ 
\hline
 \multicolumn{3}{c}{{\cellcolor[rgb]{0.957,0.957,0.957}} \textbf{\textit{Omni-modal LLMs}}} \\
\rowcolor{blue!8}{MiniCPM-o2.6-7B} &  \textbf{27.20} &  \textbf{48.20} \\
\hline
\multicolumn{3}{c}{{\cellcolor[rgb]{0.957,0.957,0.957}} \textbf{\textit{Our Models}}} \\
\rowcolor{blue!8}\model & 23.17 & 40.21  \\ \bottomrule
\end{tabular}%
}
\end{table}

\paragraph{TTS task} Finally, we evaluate \model~on Seed-TTS benchmark \cite{anastassiou2024seedttsfamilyhighqualityversatile}\footnote{https://github.com/BytedanceSpeech/seed-tts-eval}.
As shown in Table \ref{tab:tts}, our experiments demonstrate that when employing pretrained vocoders (e.g., Fishspeech1.4 or CosyVoice2.0), \model{} achieves performance comparable to its competitors (e.g., CosyVoice1.0 or CosyVoice2.0) \textbf{(Finding4)}.
\begin{table}[h!]
\centering
\caption{\textbf{Evaluation on Text-to-Speech (TTS) Benchmarks.} 
\model~demonstrated lower performance in both Mandarin (CH) and English (EN) TTS tasks. Notice: MinMo-7B's result is based on their manually filtered samples as mentioned in their report \cite{chen2025minmomultimodallargelanguage}, therefore, we do not compare with it.}\label{tab:tts}
\begin{tabular}{lcc}
\toprule
\multirow{2}{*}{\textbf{Model}} & \multicolumn{1}{c}{\textbf{CH}} & \multicolumn{1}{c}{\textbf{EN}} \\ \cmidrule{2-3}
 & \textbf{CER}$\downarrow$  & \textbf{WER}$\downarrow$ \\ \hline
\rowcolor{blue!8}CosyVoice1.0\cite{du2024cosyvoicescalablemultilingualzeroshot} & 8.64 & 3.59 \\
\rowcolor{blue!8}CosyVoice2.0\cite{du2024cosyvoice2scalablestreaming} & 4.61 & \textbf{2.43} \\
CosyVoice2.0-SFT\cite{du2024cosyvoice2scalablestreaming} & \textbf{2.06} & \underline{3.19} \\
MinMo-7B\cite{chen2025minmomultimodallargelanguage} & 2.48 & 2.90 \\
\rowcolor{blue!8}\model{} (Fishspeech1.4) & 7.55 & 13.02 \\ 
\rowcolor{blue!8}\model{} (CosyVoice2.0) & \underline{4.53} & 4.11 \\ \bottomrule
\end{tabular}%
\end{table}

Thus far, our evaluations demonstrate that incorporating the audio modality enables the model to perform audio understanding and generation without compromising its vision comprehension and reasoning capabilities. In the following section, we provide an in-depth evaluation of the vision-language alignment from the perspective of representation space.

\subsection{Representational Alignment Analysis} \label{sec:eval_embed_align}
\begin{table*}[ht!]
\centering
\caption{Kernel alignment analysis for unembedding space. The representation for each sample is the averaged token embeddings. vision+audio refers to the concatenation between audio and vision embeddings. The best two values are shown in \textbf{\textcolor{black}{bold}} and \underline{underlined}, we can observe that (1) our model without audio inputs can lead to better vision-language alignment than the backbone model on 5 out of 7 metrics, and (2) our model with audio inputs generally outperforms others on 4 out of 7 metrics.} \label{tab:eval_unembed_align}
\begin{tabular}{cccccccc} 
\toprule
Baseline & cycle knn & mutual knn & lcs knn & cka & cknna & svcca & edit knn \\ \hline
\multicolumn{8}{c}{{\cellcolor[rgb]{0.957,0.957,0.957}} \textit{\textbf{ImageNet: concepts}}} \\
\rowcolor{blue!8}Qwen2.5-VL-7B & 0.66053 & \textbf{0.04278} & \underline{1.47807} & 0.07987 & 0.02483 & 0.09168 & 0.00100 \\
Llama3.2-Vision-Instruct-11B & 0.08608 & \underline{0.04205} & \textbf{1.52788} & 0.06079 & 0.01403 & 0.11651 & 0.00061 \\
LLaVA1.5-7B & 0.48673 & 0.04164 & 1.43367 & 0.03363 & 0.01213 & 0.13553 & 0.00088 \\
LLaVA1.6-7B & 0.57173 & 0.02077 & 0.81645 & 0.08024 & 0.01577 & \underline{0.15240} & 0.00037 \\
Phi3.5-Vision-Instruct & 0.02761 & 0.01257 & 0.52355 & 0.08614 & 0.00714 & 0.12118 & 0.00019 \\
InternVL2-8B & 0.08175 & 0.01637 & 0.72495 & \underline{0.09185} & 0.00062 & 0.12148 & 0.00044\\ \hline
\rowcolor{blue!8}\model & \textbf{0.67839} & 0.03418 & 1.25284 & 0.08758 & \underline{0.02822} & 0.11963 & \underline{0.00112} \\
\rowcolor{blue!8}vision+audio & \underline{0.66432} & 0.03935 & 1.39794 & \textbf{0.09706} & \textbf{0.02833} & \textbf{0.16667} & \textbf{0.00139} \\

\toprule
\end{tabular}
\end{table*}

We formalise MLLM within the framework of an \textbf{\textit{unembedding-embedding}} architecture. In this framework, the unembedding stage is responsible for learning transformations between observations (e.g., text, vision, audio) and latent spaces through encoders, while the embedding stage captures the complex interactions among latent variables within the latent space of the hidden layers in MLLMs. Each stage serves distinct functions and yields representations with different properties. Consequently, by focusing on each stage independently, we can have a systematical evaluation of model behaviours in representation spaces. 

To assess the representational alignment among vision-language modalities at each stage, we first extract embeddings for each sample's modalities. Our evaluation focuses on two aspects: (1) whether our model inherently promotes superior vision-language alignment, and (2) whether incorporating the audio modality into the vision modality further enhances this alignment.

Quantitative evaluation of vision-language alignment is performed using kernel-alignment metrics. The kernel of a space encapsulates its underlying geometrical structure (i.e., distance metric), and the similarity between the kernels of different spaces serves as an indicator of the alignment between these spaces—in our case, across different modalities. Those metrics involve cka \citep{kornblith2019similarity}, cknna \citep{pmlr-v235-huh24a}, svcca \citep{raghu2017svcca}, and $k$-nearest neighbours (knn)-based metrics: cycle knn, mutual knn, lcs knn, and edit knn. More information for those metrics can be found in \citep{pmlr-v235-huh24a}.

\begin{figure*}[ht!]
    \centering
    \includegraphics[width=0.78\textwidth]{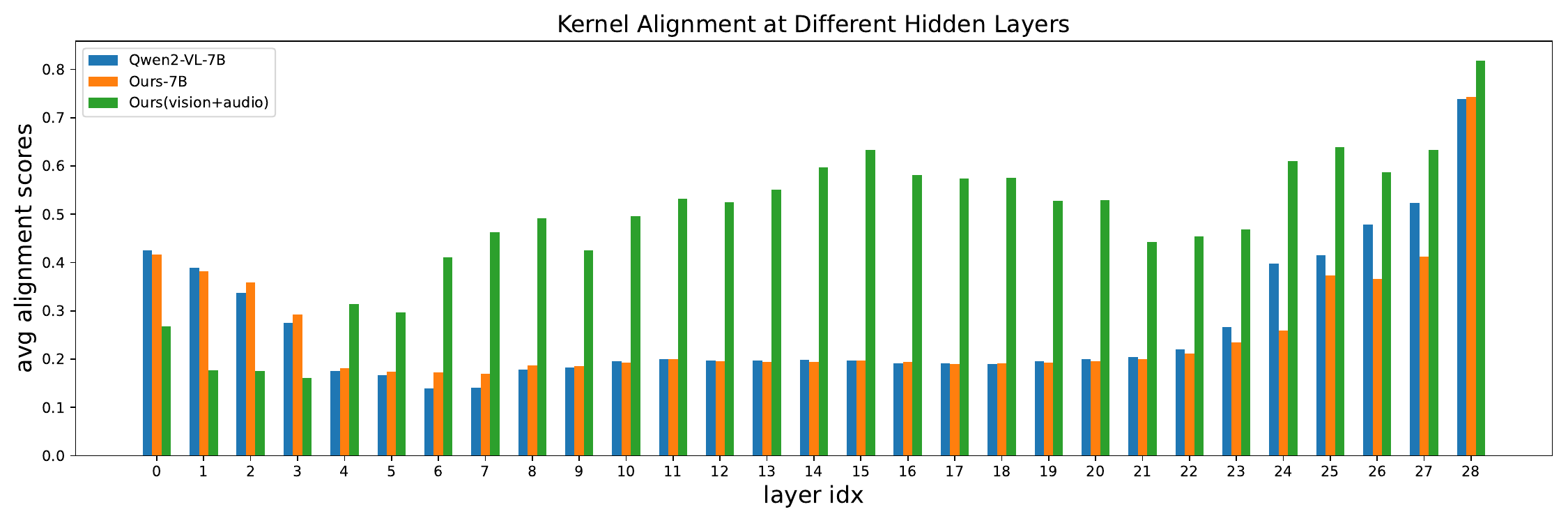}
    \caption{Averaged kernel-alignment score across different hidden layers. vision+audio: both visual and auditory modalities are concurrently fed into the model. The final fused representation is the last token in the sequence as the comprehensive summary of the integrated modalities. we can observe that incorporating audio modality (green bar) can result in better vision-language alignment at most hidden layers.}
    \label{fig:kernel_bar}
\end{figure*}
\paragraph{\textbf{Unembedding stage}} In the unembedding stage, we evaluate representational alignment using ImageNet \citep{5206848}, which offers a fine granularity in language-vision alignment, enabling a detailed assessment of representational performance. To get the audio input for each language-vision pair, we apply the pretrained Text-to-Speech LLMs, cosyvoice-300M-SFT checkpoint \citep{du2024cosyvoice}. As for the baseline, we compare our model with MLLMs after SFT, including Qwen2.5-VL 
 \citep{bai2025qwen25vltechnicalreport}, LLaMA3.2-Vision \citep{llama3.2}, LLaVA family \cite{liu2023llava}, Phi-3.5-Vision \cite{abdin2024phi}, and InternVL-8B \citep{chen2024internvl}. 

As shown in Table \ref{tab:eval_unembed_align}, we can observe that (1) our model without audio inputs can lead to better vision-language alignment than the backbone model, Qwen2.5-VL-7B, on 5 out of 7 metrics , and (2) by incorporating audio modalities, our model outperforms others on 4 out of 7 metrics \textbf{(Finding5)}. These results indirectly demonstrate that our modular formalism design concept can yield an efficient tri-modal alignment at the encoding stage.
 
 

\paragraph{\textbf{Embedding stage}} Next, we focus on the vision-language alignment at the embedding stage. We restrict our attention to the Wikipedia-caption corpus \citep{srinivasan2021wit}. 
We present a comparative analysis between Qwen2.5-VL-7B and our model across all hidden layers. We feed both image and its audio into the model and consider the last token embedding as the final fused representation. As illustrated in Figure \ref{fig:kernel_bar}, the incorporation of the audio modality significantly enhances vision-language alignment, particularly in the intermediate hidden layers \textbf{(Finding6)}. This finding underscores the efficacy of our training strategy, whereby pretraining exclusively on audio data yields a robust tri-modal alignment.

\section{Conclusion and Future Work} \label{sec:conclusion}

In this work, we propose a comprehensive, industry-level omni-modal LLM pipeline that comprises a modularised end-to-end framework, a lightweight training strategy, and an audio data synthesis pipeline to overcome the current challenges, including limited accessibility to tri-modal datasets, higher computational resource requirements, and the complexity of aligning features across three modalities. To this end, we propose an industry-level omni-perceive and -interactive tri-modal LLM, \textbf{\model{}}. 

Extensive experiments validate the usefulness of our proposed pipeline and \model. In the visual understanding task, \model~exhibits superior performance compared with its backbone model - Qwen2.5-VL-7B, validating the efficiency of our training strategy. In the Text-to-Speech task, based on pretrained vocoder (e.g., Fishspeech1.4 or CosyVoice2.0), \model{} is comparable to its backbone vocoder on Seed-TTS benchmark. In our real-world ASR testset, \model{} achieves outstanding performance, indicating its robustness in real scenarios. 
In the future, we intend to link \model{} with vision generative models under our modularised framework to allow the model to seamlessly generate high-quality, contextually coherent visual outputs. Integrating the Artificial Intelligence-Generated Content (AIGC) capabilities into the current model can further broaden its applicability in creative and industrial domains, such as Embodied Intelligence.
\bibliographystyle{ACM-Reference-Format}
\bibliography{sample-base}


\begin{thebibliography}{76}


\ifx \showCODEN    \undefined \def \showCODEN     #1{\unskip}     \fi
\ifx \showISBNx    \undefined \def \showISBNx     #1{\unskip}     \fi
\ifx \showISBNxiii \undefined \def \showISBNxiii  #1{\unskip}     \fi
\ifx \showISSN     \undefined \def \showISSN      #1{\unskip}     \fi
\ifx \showLCCN     \undefined \def \showLCCN      #1{\unskip}     \fi
\ifx \shownote     \undefined \def \shownote      #1{#1}          \fi
\ifx \showarticletitle \undefined \def \showarticletitle #1{#1}   \fi
\ifx \showURL      \undefined \def \showURL       {\relax}        \fi
\providecommand\bibfield[2]{#2}
\providecommand\bibinfo[2]{#2}
\providecommand\natexlab[1]{#1}
\providecommand\showeprint[2][]{arXiv:#2}

\bibitem[Abdin et~al\mbox{.}(2024)]%
        {abdin2024phi}
\bibfield{author}{\bibinfo{person}{Marah Abdin}, \bibinfo{person}{Jyoti Aneja}, \bibinfo{person}{Hany Awadalla}, \bibinfo{person}{Ahmed Awadallah}, \bibinfo{person}{Ammar~Ahmad Awan}, \bibinfo{person}{Nguyen Bach}, \bibinfo{person}{Amit Bahree}, \bibinfo{person}{Arash Bakhtiari}, \bibinfo{person}{Jianmin Bao}, \bibinfo{person}{Harkirat Behl}, {et~al\mbox{.}}} \bibinfo{year}{2024}\natexlab{}.
\newblock \showarticletitle{Phi-3 technical report: A highly capable language model locally on your phone}.
\newblock \bibinfo{journal}{\emph{arXiv preprint arXiv:2404.14219}} (\bibinfo{year}{2024}).
\newblock


\bibitem[Anastassiou et~al\mbox{.}(2024)]%
        {anastassiou2024seedttsfamilyhighqualityversatile}
\bibfield{author}{\bibinfo{person}{Philip Anastassiou}, \bibinfo{person}{Jiawei Chen}, \bibinfo{person}{Jitong Chen}, \bibinfo{person}{Yuanzhe Chen}, \bibinfo{person}{Zhuo Chen}, \bibinfo{person}{Ziyi Chen}, \bibinfo{person}{Jian Cong}, \bibinfo{person}{Lelai Deng}, \bibinfo{person}{Chuang Ding}, \bibinfo{person}{Lu Gao}, \bibinfo{person}{Mingqing Gong}, \bibinfo{person}{Peisong Huang}, \bibinfo{person}{Qingqing Huang}, \bibinfo{person}{Zhiying Huang}, \bibinfo{person}{Yuanyuan Huo}, \bibinfo{person}{Dongya Jia}, \bibinfo{person}{Chumin Li}, \bibinfo{person}{Feiya Li}, \bibinfo{person}{Hui Li}, \bibinfo{person}{Jiaxin Li}, \bibinfo{person}{Xiaoyang Li}, \bibinfo{person}{Xingxing Li}, \bibinfo{person}{Lin Liu}, \bibinfo{person}{Shouda Liu}, \bibinfo{person}{Sichao Liu}, \bibinfo{person}{Xudong Liu}, \bibinfo{person}{Yuchen Liu}, \bibinfo{person}{Zhengxi Liu}, \bibinfo{person}{Lu Lu}, \bibinfo{person}{Junjie Pan}, \bibinfo{person}{Xin Wang}, \bibinfo{person}{Yuping Wang}, \bibinfo{person}{Yuxuan Wang},
  \bibinfo{person}{Zhen Wei}, \bibinfo{person}{Jian Wu}, \bibinfo{person}{Chao Yao}, \bibinfo{person}{Yifeng Yang}, \bibinfo{person}{Yuanhao Yi}, \bibinfo{person}{Junteng Zhang}, \bibinfo{person}{Qidi Zhang}, \bibinfo{person}{Shuo Zhang}, \bibinfo{person}{Wenjie Zhang}, \bibinfo{person}{Yang Zhang}, \bibinfo{person}{Zilin Zhao}, \bibinfo{person}{Dejian Zhong}, {and} \bibinfo{person}{Xiaobin Zhuang}.} \bibinfo{year}{2024}\natexlab{}.
\newblock \bibinfo{title}{Seed-TTS: A Family of High-Quality Versatile Speech Generation Models}.
\newblock
\showeprint[arxiv]{2406.02430}~[eess.AS]
\urldef\tempurl%
\url{https://arxiv.org/abs/2406.02430}
\showURL{%
\tempurl}


\bibitem[Ardila et~al\mbox{.}(2019)]%
        {ardila2019common}
\bibfield{author}{\bibinfo{person}{Rosana Ardila}, \bibinfo{person}{Megan Branson}, \bibinfo{person}{Kelly Davis}, \bibinfo{person}{Michael Henretty}, \bibinfo{person}{Michael Kohler}, \bibinfo{person}{Josh Meyer}, \bibinfo{person}{Reuben Morais}, \bibinfo{person}{Lindsay Saunders}, \bibinfo{person}{Francis~M Tyers}, {and} \bibinfo{person}{Gregor Weber}.} \bibinfo{year}{2019}\natexlab{}.
\newblock \showarticletitle{Common voice: A massively-multilingual speech corpus}.
\newblock \bibinfo{journal}{\emph{arXiv preprint arXiv:1912.06670}} (\bibinfo{year}{2019}).
\newblock


\bibitem[Bai et~al\mbox{.}(2025)]%
        {bai2025qwen25vltechnicalreport}
\bibfield{author}{\bibinfo{person}{Shuai Bai}, \bibinfo{person}{Keqin Chen}, \bibinfo{person}{Xuejing Liu}, \bibinfo{person}{Jialin Wang}, \bibinfo{person}{Wenbin Ge}, \bibinfo{person}{Sibo Song}, \bibinfo{person}{Kai Dang}, \bibinfo{person}{Peng Wang}, \bibinfo{person}{Shijie Wang}, \bibinfo{person}{Jun Tang}, \bibinfo{person}{Humen Zhong}, \bibinfo{person}{Yuanzhi Zhu}, \bibinfo{person}{Mingkun Yang}, \bibinfo{person}{Zhaohai Li}, \bibinfo{person}{Jianqiang Wan}, \bibinfo{person}{Pengfei Wang}, \bibinfo{person}{Wei Ding}, \bibinfo{person}{Zheren Fu}, \bibinfo{person}{Yiheng Xu}, \bibinfo{person}{Jiabo Ye}, \bibinfo{person}{Xi Zhang}, \bibinfo{person}{Tianbao Xie}, \bibinfo{person}{Zesen Cheng}, \bibinfo{person}{Hang Zhang}, \bibinfo{person}{Zhibo Yang}, \bibinfo{person}{Haiyang Xu}, {and} \bibinfo{person}{Junyang Lin}.} \bibinfo{year}{2025}\natexlab{}.
\newblock \bibinfo{title}{Qwen2.5-VL Technical Report}.
\newblock
\showeprint[arxiv]{2502.13923}~[cs.CV]
\urldef\tempurl%
\url{https://arxiv.org/abs/2502.13923}
\showURL{%
\tempurl}


\bibitem[Bu et~al\mbox{.}(2017)]%
        {bu2017aishell1opensourcemandarinspeech}
\bibfield{author}{\bibinfo{person}{Hui Bu}, \bibinfo{person}{Jiayu Du}, \bibinfo{person}{Xingyu Na}, \bibinfo{person}{Bengu Wu}, {and} \bibinfo{person}{Hao Zheng}.} \bibinfo{year}{2017}\natexlab{}.
\newblock \bibinfo{title}{AISHELL-1: An Open-Source Mandarin Speech Corpus and A Speech Recognition Baseline}.
\newblock
\showeprint[arxiv]{1709.05522}~[cs.CL]
\urldef\tempurl%
\url{https://arxiv.org/abs/1709.05522}
\showURL{%
\tempurl}


\bibitem[Chen et~al\mbox{.}(2025c)]%
        {chen2025emovaempoweringlanguagemodels}
\bibfield{author}{\bibinfo{person}{Kai Chen}, \bibinfo{person}{Yunhao Gou}, \bibinfo{person}{Runhui Huang}, \bibinfo{person}{Zhili Liu}, \bibinfo{person}{Daxin Tan}, \bibinfo{person}{Jing Xu}, \bibinfo{person}{Chunwei Wang}, \bibinfo{person}{Yi Zhu}, \bibinfo{person}{Yihan Zeng}, \bibinfo{person}{Kuo Yang}, \bibinfo{person}{Dingdong Wang}, \bibinfo{person}{Kun Xiang}, \bibinfo{person}{Haoyuan Li}, \bibinfo{person}{Haoli Bai}, \bibinfo{person}{Jianhua Han}, \bibinfo{person}{Xiaohui Li}, \bibinfo{person}{Weike Jin}, \bibinfo{person}{Nian Xie}, \bibinfo{person}{Yu Zhang}, \bibinfo{person}{James~T. Kwok}, \bibinfo{person}{Hengshuang Zhao}, \bibinfo{person}{Xiaodan Liang}, \bibinfo{person}{Dit-Yan Yeung}, \bibinfo{person}{Xiao Chen}, \bibinfo{person}{Zhenguo Li}, \bibinfo{person}{Wei Zhang}, \bibinfo{person}{Qun Liu}, \bibinfo{person}{Jun Yao}, \bibinfo{person}{Lanqing Hong}, \bibinfo{person}{Lu Hou}, {and} \bibinfo{person}{Hang Xu}.} \bibinfo{year}{2025}\natexlab{c}.
\newblock \bibinfo{title}{EMOVA: Empowering Language Models to See, Hear and Speak with Vivid Emotions}.
\newblock
\showeprint[arxiv]{2409.18042}~[cs.CV]
\urldef\tempurl%
\url{https://arxiv.org/abs/2409.18042}
\showURL{%
\tempurl}


\bibitem[Chen et~al\mbox{.}(2025a)]%
        {chen2025minmo}
\bibfield{author}{\bibinfo{person}{Qian Chen}, \bibinfo{person}{Yafeng Chen}, \bibinfo{person}{Yanni Chen}, \bibinfo{person}{Mengzhe Chen}, \bibinfo{person}{Yingda Chen}, \bibinfo{person}{Chong Deng}, \bibinfo{person}{Zhihao Du}, \bibinfo{person}{Ruize Gao}, \bibinfo{person}{Changfeng Gao}, \bibinfo{person}{Zhifu Gao}, {et~al\mbox{.}}} \bibinfo{year}{2025}\natexlab{a}.
\newblock \showarticletitle{MinMo: A Multimodal Large Language Model for Seamless Voice Interaction}.
\newblock \bibinfo{journal}{\emph{arXiv preprint arXiv:2501.06282}} (\bibinfo{year}{2025}).
\newblock


\bibitem[Chen et~al\mbox{.}(2025b)]%
        {chen2025minmomultimodallargelanguage}
\bibfield{author}{\bibinfo{person}{Qian Chen}, \bibinfo{person}{Yafeng Chen}, \bibinfo{person}{Yanni Chen}, \bibinfo{person}{Mengzhe Chen}, \bibinfo{person}{Yingda Chen}, \bibinfo{person}{Chong Deng}, \bibinfo{person}{Zhihao Du}, \bibinfo{person}{Ruize Gao}, \bibinfo{person}{Changfeng Gao}, \bibinfo{person}{Zhifu Gao}, \bibinfo{person}{Yabin Li}, \bibinfo{person}{Xiang Lv}, \bibinfo{person}{Jiaqing Liu}, \bibinfo{person}{Haoneng Luo}, \bibinfo{person}{Bin Ma}, \bibinfo{person}{Chongjia Ni}, \bibinfo{person}{Xian Shi}, \bibinfo{person}{Jialong Tang}, \bibinfo{person}{Hui Wang}, \bibinfo{person}{Hao Wang}, \bibinfo{person}{Wen Wang}, \bibinfo{person}{Yuxuan Wang}, \bibinfo{person}{Yunlan Xu}, \bibinfo{person}{Fan Yu}, \bibinfo{person}{Zhijie Yan}, \bibinfo{person}{Yexin Yang}, \bibinfo{person}{Baosong Yang}, \bibinfo{person}{Xian Yang}, \bibinfo{person}{Guanrou Yang}, \bibinfo{person}{Tianyu Zhao}, \bibinfo{person}{Qinglin Zhang}, \bibinfo{person}{Shiliang Zhang}, \bibinfo{person}{Nan Zhao},
  \bibinfo{person}{Pei Zhang}, \bibinfo{person}{Chong Zhang}, {and} \bibinfo{person}{Jinren Zhou}.} \bibinfo{year}{2025}\natexlab{b}.
\newblock \bibinfo{title}{MinMo: A Multimodal Large Language Model for Seamless Voice Interaction}.
\newblock
\showeprint[arxiv]{2501.06282}~[cs.CL]
\urldef\tempurl%
\url{https://arxiv.org/abs/2501.06282}
\showURL{%
\tempurl}


\bibitem[Chen et~al\mbox{.}(2024a)]%
        {chen2024fargpt4vclosinggap}
\bibfield{author}{\bibinfo{person}{Zhe Chen}, \bibinfo{person}{Weiyun Wang}, \bibinfo{person}{Hao Tian}, \bibinfo{person}{Shenglong Ye}, \bibinfo{person}{Zhangwei Gao}, \bibinfo{person}{Erfei Cui}, \bibinfo{person}{Wenwen Tong}, \bibinfo{person}{Kongzhi Hu}, \bibinfo{person}{Jiapeng Luo}, \bibinfo{person}{Zheng Ma}, \bibinfo{person}{Ji Ma}, \bibinfo{person}{Jiaqi Wang}, \bibinfo{person}{Xiaoyi Dong}, \bibinfo{person}{Hang Yan}, \bibinfo{person}{Hewei Guo}, \bibinfo{person}{Conghui He}, \bibinfo{person}{Botian Shi}, \bibinfo{person}{Zhenjiang Jin}, \bibinfo{person}{Chao Xu}, \bibinfo{person}{Bin Wang}, \bibinfo{person}{Xingjian Wei}, \bibinfo{person}{Wei Li}, \bibinfo{person}{Wenjian Zhang}, \bibinfo{person}{Bo Zhang}, \bibinfo{person}{Pinlong Cai}, \bibinfo{person}{Licheng Wen}, \bibinfo{person}{Xiangchao Yan}, \bibinfo{person}{Min Dou}, \bibinfo{person}{Lewei Lu}, \bibinfo{person}{Xizhou Zhu}, \bibinfo{person}{Tong Lu}, \bibinfo{person}{Dahua Lin}, \bibinfo{person}{Yu Qiao}, \bibinfo{person}{Jifeng Dai},
  {and} \bibinfo{person}{Wenhai Wang}.} \bibinfo{year}{2024}\natexlab{a}.
\newblock \bibinfo{title}{How Far Are We to GPT-4V? Closing the Gap to Commercial Multimodal Models with Open-Source Suites}.
\newblock
\showeprint[arxiv]{2404.16821}~[cs.CV]
\urldef\tempurl%
\url{https://arxiv.org/abs/2404.16821}
\showURL{%
\tempurl}


\bibitem[Chen et~al\mbox{.}(2024b)]%
        {chen2024internvl}
\bibfield{author}{\bibinfo{person}{Zhe Chen}, \bibinfo{person}{Jiannan Wu}, \bibinfo{person}{Wenhai Wang}, \bibinfo{person}{Weijie Su}, \bibinfo{person}{Guo Chen}, \bibinfo{person}{Sen Xing}, \bibinfo{person}{Muyan Zhong}, \bibinfo{person}{Qinglong Zhang}, \bibinfo{person}{Xizhou Zhu}, \bibinfo{person}{Lewei Lu}, {et~al\mbox{.}}} \bibinfo{year}{2024}\natexlab{b}.
\newblock \showarticletitle{Internvl: Scaling up vision foundation models and aligning for generic visual-linguistic tasks}. In \bibinfo{booktitle}{\emph{Proceedings of the IEEE/CVF Conference on Computer Vision and Pattern Recognition}}. \bibinfo{pages}{24185--24198}.
\newblock


\bibitem[Chu et~al\mbox{.}(2024a)]%
        {chu2024qwen2}
\bibfield{author}{\bibinfo{person}{Yunfei Chu}, \bibinfo{person}{Jin Xu}, \bibinfo{person}{Qian Yang}, \bibinfo{person}{Haojie Wei}, \bibinfo{person}{Xipin Wei}, \bibinfo{person}{Zhifang Guo}, \bibinfo{person}{Yichong Leng}, \bibinfo{person}{Yuanjun Lv}, \bibinfo{person}{Jinzheng He}, \bibinfo{person}{Junyang Lin}, {et~al\mbox{.}}} \bibinfo{year}{2024}\natexlab{a}.
\newblock \showarticletitle{Qwen2-audio technical report}.
\newblock \bibinfo{journal}{\emph{arXiv preprint arXiv:2407.10759}} (\bibinfo{year}{2024}).
\newblock


\bibitem[Chu et~al\mbox{.}(2024b)]%
        {chu2024qwen2audiotechnicalreport}
\bibfield{author}{\bibinfo{person}{Yunfei Chu}, \bibinfo{person}{Jin Xu}, \bibinfo{person}{Qian Yang}, \bibinfo{person}{Haojie Wei}, \bibinfo{person}{Xipin Wei}, \bibinfo{person}{Zhifang Guo}, \bibinfo{person}{Yichong Leng}, \bibinfo{person}{Yuanjun Lv}, \bibinfo{person}{Jinzheng He}, \bibinfo{person}{Junyang Lin}, \bibinfo{person}{Chang Zhou}, {and} \bibinfo{person}{Jingren Zhou}.} \bibinfo{year}{2024}\natexlab{b}.
\newblock \bibinfo{title}{Qwen2-Audio Technical Report}.
\newblock
\showeprint[arxiv]{2407.10759}~[eess.AS]
\urldef\tempurl%
\url{https://arxiv.org/abs/2407.10759}
\showURL{%
\tempurl}


\bibitem[Conneau et~al\mbox{.}(2023)]%
        {conneau2023fleurs}
\bibfield{author}{\bibinfo{person}{Alexis Conneau}, \bibinfo{person}{Min Ma}, \bibinfo{person}{Simran Khanuja}, \bibinfo{person}{Yu Zhang}, \bibinfo{person}{Vera Axelrod}, \bibinfo{person}{Siddharth Dalmia}, \bibinfo{person}{Jason Riesa}, \bibinfo{person}{Clara Rivera}, {and} \bibinfo{person}{Ankur Bapna}.} \bibinfo{year}{2023}\natexlab{}.
\newblock \showarticletitle{Fleurs: Few-shot learning evaluation of universal representations of speech}. In \bibinfo{booktitle}{\emph{2022 IEEE Spoken Language Technology Workshop (SLT)}}. IEEE, \bibinfo{pages}{798--805}.
\newblock


\bibitem[Copet et~al\mbox{.}(2023)]%
        {copet2023simple}
\bibfield{author}{\bibinfo{person}{Jade Copet}, \bibinfo{person}{Felix Kreuk}, \bibinfo{person}{Itai Gat}, \bibinfo{person}{Tal Remez}, \bibinfo{person}{David Kant}, \bibinfo{person}{Gabriel Synnaeve}, \bibinfo{person}{Yossi Adi}, {and} \bibinfo{person}{Alexandre D{\'e}fossez}.} \bibinfo{year}{2023}\natexlab{}.
\newblock \showarticletitle{Simple and controllable music generation}.
\newblock \bibinfo{journal}{\emph{Advances in Neural Information Processing Systems}}  \bibinfo{volume}{36} (\bibinfo{year}{2023}).
\newblock


\bibitem[Dai et~al\mbox{.}(2022)]%
        {dai2022one}
\bibfield{author}{\bibinfo{person}{Yong Dai}, \bibinfo{person}{Duyu Tang}, \bibinfo{person}{Liangxin Liu}, \bibinfo{person}{Minghuan Tan}, \bibinfo{person}{Cong Zhou}, \bibinfo{person}{Jingquan Wang}, \bibinfo{person}{Zhangyin Feng}, \bibinfo{person}{Fan Zhang}, \bibinfo{person}{Xueyu Hu}, {and} \bibinfo{person}{Shuming Shi}.} \bibinfo{year}{2022}\natexlab{}.
\newblock \showarticletitle{One model, multiple modalities: A sparsely activated approach for text, sound, image, video and code}.
\newblock \bibinfo{journal}{\emph{arXiv preprint arXiv:2205.06126}} (\bibinfo{year}{2022}).
\newblock


\bibitem[D{\'e}fossez et~al\mbox{.}(2024)]%
        {defossez2024moshi}
\bibfield{author}{\bibinfo{person}{Alexandre D{\'e}fossez}, \bibinfo{person}{Laurent Mazar{\'e}}, \bibinfo{person}{Manu Orsini}, \bibinfo{person}{Am{\'e}lie Royer}, \bibinfo{person}{Patrick P{\'e}rez}, \bibinfo{person}{Herv{\'e} J{\'e}gou}, \bibinfo{person}{Edouard Grave}, {and} \bibinfo{person}{Neil Zeghidour}.} \bibinfo{year}{2024}\natexlab{}.
\newblock \showarticletitle{Moshi: a speech-text foundation model for real-time dialogue}.
\newblock \bibinfo{journal}{\emph{arXiv preprint arXiv:2410.00037}} (\bibinfo{year}{2024}).
\newblock


\bibitem[Deng et~al\mbox{.}(2009)]%
        {5206848}
\bibfield{author}{\bibinfo{person}{Jia Deng}, \bibinfo{person}{Wei Dong}, \bibinfo{person}{Richard Socher}, \bibinfo{person}{Li-Jia Li}, \bibinfo{person}{Kai Li}, {and} \bibinfo{person}{Li Fei-Fei}.} \bibinfo{year}{2009}\natexlab{}.
\newblock \showarticletitle{ImageNet: A large-scale hierarchical image database}. In \bibinfo{booktitle}{\emph{2009 IEEE Conference on Computer Vision and Pattern Recognition}}. \bibinfo{pages}{248--255}.
\newblock
\href{https://doi.org/10.1109/CVPR.2009.5206848}{doi:\nolinkurl{10.1109/CVPR.2009.5206848}}


\bibitem[Dong et~al\mbox{.}(2024)]%
        {dong2024internlmxcomposer2masteringfreeformtextimage}
\bibfield{author}{\bibinfo{person}{Xiaoyi Dong}, \bibinfo{person}{Pan Zhang}, \bibinfo{person}{Yuhang Zang}, \bibinfo{person}{Yuhang Cao}, \bibinfo{person}{Bin Wang}, \bibinfo{person}{Linke Ouyang}, \bibinfo{person}{Xilin Wei}, \bibinfo{person}{Songyang Zhang}, \bibinfo{person}{Haodong Duan}, \bibinfo{person}{Maosong Cao}, \bibinfo{person}{Wenwei Zhang}, \bibinfo{person}{Yining Li}, \bibinfo{person}{Hang Yan}, \bibinfo{person}{Yang Gao}, \bibinfo{person}{Xinyue Zhang}, \bibinfo{person}{Wei Li}, \bibinfo{person}{Jingwen Li}, \bibinfo{person}{Kai Chen}, \bibinfo{person}{Conghui He}, \bibinfo{person}{Xingcheng Zhang}, \bibinfo{person}{Yu Qiao}, \bibinfo{person}{Dahua Lin}, {and} \bibinfo{person}{Jiaqi Wang}.} \bibinfo{year}{2024}\natexlab{}.
\newblock \bibinfo{title}{InternLM-XComposer2: Mastering Free-form Text-Image Composition and Comprehension in Vision-Language Large Model}.
\newblock
\showeprint[arxiv]{2401.16420}~[cs.CV]
\urldef\tempurl%
\url{https://arxiv.org/abs/2401.16420}
\showURL{%
\tempurl}


\bibitem[Du et~al\mbox{.}(2018a)]%
        {du2018aishell}
\bibfield{author}{\bibinfo{person}{Jiayu Du}, \bibinfo{person}{Xingyu Na}, \bibinfo{person}{Xuechen Liu}, {and} \bibinfo{person}{Hui Bu}.} \bibinfo{year}{2018}\natexlab{a}.
\newblock \showarticletitle{Aishell-2: Transforming mandarin asr research into industrial scale}.
\newblock \bibinfo{journal}{\emph{arXiv preprint arXiv:1808.10583}} (\bibinfo{year}{2018}).
\newblock


\bibitem[Du et~al\mbox{.}(2018b)]%
        {du2018aishell2transformingmandarinasr}
\bibfield{author}{\bibinfo{person}{Jiayu Du}, \bibinfo{person}{Xingyu Na}, \bibinfo{person}{Xuechen Liu}, {and} \bibinfo{person}{Hui Bu}.} \bibinfo{year}{2018}\natexlab{b}.
\newblock \bibinfo{title}{AISHELL-2: Transforming Mandarin ASR Research Into Industrial Scale}.
\newblock
\showeprint[arxiv]{1808.10583}~[cs.CL]
\urldef\tempurl%
\url{https://arxiv.org/abs/1808.10583}
\showURL{%
\tempurl}


\bibitem[Du et~al\mbox{.}(2024a)]%
        {du2024cosyvoice}
\bibfield{author}{\bibinfo{person}{Zhihao Du}, \bibinfo{person}{Qian Chen}, \bibinfo{person}{Shiliang Zhang}, \bibinfo{person}{Kai Hu}, \bibinfo{person}{Heng Lu}, \bibinfo{person}{Yexin Yang}, \bibinfo{person}{Hangrui Hu}, \bibinfo{person}{Siqi Zheng}, \bibinfo{person}{Yue Gu}, \bibinfo{person}{Ziyang Ma}, {et~al\mbox{.}}} \bibinfo{year}{2024}\natexlab{a}.
\newblock \showarticletitle{Cosyvoice: A scalable multilingual zero-shot text-to-speech synthesizer based on supervised semantic tokens}.
\newblock \bibinfo{journal}{\emph{arXiv preprint arXiv:2407.05407}} (\bibinfo{year}{2024}).
\newblock


\bibitem[Du et~al\mbox{.}(2024b)]%
        {du2024cosyvoicescalablemultilingualzeroshot}
\bibfield{author}{\bibinfo{person}{Zhihao Du}, \bibinfo{person}{Qian Chen}, \bibinfo{person}{Shiliang Zhang}, \bibinfo{person}{Kai Hu}, \bibinfo{person}{Heng Lu}, \bibinfo{person}{Yexin Yang}, \bibinfo{person}{Hangrui Hu}, \bibinfo{person}{Siqi Zheng}, \bibinfo{person}{Yue Gu}, \bibinfo{person}{Ziyang Ma}, \bibinfo{person}{Zhifu Gao}, {and} \bibinfo{person}{Zhijie Yan}.} \bibinfo{year}{2024}\natexlab{b}.
\newblock \bibinfo{title}{CosyVoice: A Scalable Multilingual Zero-shot Text-to-speech Synthesizer based on Supervised Semantic Tokens}.
\newblock
\showeprint[arxiv]{2407.05407}~[cs.SD]
\urldef\tempurl%
\url{https://arxiv.org/abs/2407.05407}
\showURL{%
\tempurl}


\bibitem[Du et~al\mbox{.}(2024c)]%
        {du2024cosyvoice2scalablestreaming}
\bibfield{author}{\bibinfo{person}{Zhihao Du}, \bibinfo{person}{Yuxuan Wang}, \bibinfo{person}{Qian Chen}, \bibinfo{person}{Xian Shi}, \bibinfo{person}{Xiang Lv}, \bibinfo{person}{Tianyu Zhao}, \bibinfo{person}{Zhifu Gao}, \bibinfo{person}{Yexin Yang}, \bibinfo{person}{Changfeng Gao}, \bibinfo{person}{Hui Wang}, \bibinfo{person}{Fan Yu}, \bibinfo{person}{Huadai Liu}, \bibinfo{person}{Zhengyan Sheng}, \bibinfo{person}{Yue Gu}, \bibinfo{person}{Chong Deng}, \bibinfo{person}{Wen Wang}, \bibinfo{person}{Shiliang Zhang}, \bibinfo{person}{Zhijie Yan}, {and} \bibinfo{person}{Jingren Zhou}.} \bibinfo{year}{2024}\natexlab{c}.
\newblock \bibinfo{title}{CosyVoice 2: Scalable Streaming Speech Synthesis with Large Language Models}.
\newblock
\showeprint[arxiv]{2412.10117}~[cs.SD]
\urldef\tempurl%
\url{https://arxiv.org/abs/2412.10117}
\showURL{%
\tempurl}


\bibitem[Fang et~al\mbox{.}(2024a)]%
        {fang2024llama}
\bibfield{author}{\bibinfo{person}{Qingkai Fang}, \bibinfo{person}{Shoutao Guo}, \bibinfo{person}{Yan Zhou}, \bibinfo{person}{Zhengrui Ma}, \bibinfo{person}{Shaolei Zhang}, {and} \bibinfo{person}{Yang Feng}.} \bibinfo{year}{2024}\natexlab{a}.
\newblock \showarticletitle{Llama-omni: Seamless speech interaction with large language models}.
\newblock \bibinfo{journal}{\emph{arXiv preprint arXiv:2409.06666}} (\bibinfo{year}{2024}).
\newblock


\bibitem[Fang et~al\mbox{.}(2024b)]%
        {fang2024llamaomniseamlessspeechinteraction}
\bibfield{author}{\bibinfo{person}{Qingkai Fang}, \bibinfo{person}{Shoutao Guo}, \bibinfo{person}{Yan Zhou}, \bibinfo{person}{Zhengrui Ma}, \bibinfo{person}{Shaolei Zhang}, {and} \bibinfo{person}{Yang Feng}.} \bibinfo{year}{2024}\natexlab{b}.
\newblock \bibinfo{title}{LLaMA-Omni: Seamless Speech Interaction with Large Language Models}.
\newblock
\showeprint[arxiv]{2409.06666}~[cs.CL]
\urldef\tempurl%
\url{https://arxiv.org/abs/2409.06666}
\showURL{%
\tempurl}


\bibitem[Fu et~al\mbox{.}(2024a)]%
        {fu2024mmecomprehensiveevaluationbenchmark}
\bibfield{author}{\bibinfo{person}{Chaoyou Fu}, \bibinfo{person}{Peixian Chen}, \bibinfo{person}{Yunhang Shen}, \bibinfo{person}{Yulei Qin}, \bibinfo{person}{Mengdan Zhang}, \bibinfo{person}{Xu Lin}, \bibinfo{person}{Jinrui Yang}, \bibinfo{person}{Xiawu Zheng}, \bibinfo{person}{Ke Li}, \bibinfo{person}{Xing Sun}, \bibinfo{person}{Yunsheng Wu}, {and} \bibinfo{person}{Rongrong Ji}.} \bibinfo{year}{2024}\natexlab{a}.
\newblock \bibinfo{title}{MME: A Comprehensive Evaluation Benchmark for Multimodal Large Language Models}.
\newblock
\showeprint[arxiv]{2306.13394}~[cs.CV]
\urldef\tempurl%
\url{https://arxiv.org/abs/2306.13394}
\showURL{%
\tempurl}


\bibitem[Fu et~al\mbox{.}(2024b)]%
        {fu2024videommefirstevercomprehensiveevaluation}
\bibfield{author}{\bibinfo{person}{Chaoyou Fu}, \bibinfo{person}{Yuhan Dai}, \bibinfo{person}{Yongdong Luo}, \bibinfo{person}{Lei Li}, \bibinfo{person}{Shuhuai Ren}, \bibinfo{person}{Renrui Zhang}, \bibinfo{person}{Zihan Wang}, \bibinfo{person}{Chenyu Zhou}, \bibinfo{person}{Yunhang Shen}, \bibinfo{person}{Mengdan Zhang}, \bibinfo{person}{Peixian Chen}, \bibinfo{person}{Yanwei Li}, \bibinfo{person}{Shaohui Lin}, \bibinfo{person}{Sirui Zhao}, \bibinfo{person}{Ke Li}, \bibinfo{person}{Tong Xu}, \bibinfo{person}{Xiawu Zheng}, \bibinfo{person}{Enhong Chen}, \bibinfo{person}{Rongrong Ji}, {and} \bibinfo{person}{Xing Sun}.} \bibinfo{year}{2024}\natexlab{b}.
\newblock \bibinfo{title}{Video-MME: The First-Ever Comprehensive Evaluation Benchmark of Multi-modal LLMs in Video Analysis}.
\newblock
\showeprint[arxiv]{2405.21075}~[cs.CV]
\urldef\tempurl%
\url{https://arxiv.org/abs/2405.21075}
\showURL{%
\tempurl}


\bibitem[Fu et~al\mbox{.}(2024c)]%
        {fu2024vita}
\bibfield{author}{\bibinfo{person}{Chaoyou Fu}, \bibinfo{person}{Haojia Lin}, \bibinfo{person}{Zuwei Long}, \bibinfo{person}{Yunhang Shen}, \bibinfo{person}{Meng Zhao}, \bibinfo{person}{Yifan Zhang}, \bibinfo{person}{Shaoqi Dong}, \bibinfo{person}{Xiong Wang}, \bibinfo{person}{Di Yin}, \bibinfo{person}{Long Ma}, {et~al\mbox{.}}} \bibinfo{year}{2024}\natexlab{c}.
\newblock \showarticletitle{Vita: Towards open-source interactive omni multimodal llm}.
\newblock \bibinfo{journal}{\emph{arXiv preprint arXiv:2408.05211}} (\bibinfo{year}{2024}).
\newblock


\bibitem[Fu et~al\mbox{.}(2025a)]%
        {fu2025vita15gpt4olevelrealtime}
\bibfield{author}{\bibinfo{person}{Chaoyou Fu}, \bibinfo{person}{Haojia Lin}, \bibinfo{person}{Xiong Wang}, \bibinfo{person}{Yi-Fan Zhang}, \bibinfo{person}{Yunhang Shen}, \bibinfo{person}{Xiaoyu Liu}, \bibinfo{person}{Haoyu Cao}, \bibinfo{person}{Zuwei Long}, \bibinfo{person}{Heting Gao}, \bibinfo{person}{Ke Li}, \bibinfo{person}{Long Ma}, \bibinfo{person}{Xiawu Zheng}, \bibinfo{person}{Rongrong Ji}, \bibinfo{person}{Xing Sun}, \bibinfo{person}{Caifeng Shan}, {and} \bibinfo{person}{Ran He}.} \bibinfo{year}{2025}\natexlab{a}.
\newblock \bibinfo{title}{VITA-1.5: Towards GPT-4o Level Real-Time Vision and Speech Interaction}.
\newblock
\showeprint[arxiv]{2501.01957}~[cs.CV]
\urldef\tempurl%
\url{https://arxiv.org/abs/2501.01957}
\showURL{%
\tempurl}


\bibitem[Fu et~al\mbox{.}(2025b)]%
        {fu2025vita}
\bibfield{author}{\bibinfo{person}{Chaoyou Fu}, \bibinfo{person}{Haojia Lin}, \bibinfo{person}{Xiong Wang}, \bibinfo{person}{Yi-Fan Zhang}, \bibinfo{person}{Yunhang Shen}, \bibinfo{person}{Xiaoyu Liu}, \bibinfo{person}{Yangze Li}, \bibinfo{person}{Zuwei Long}, \bibinfo{person}{Heting Gao}, \bibinfo{person}{Ke Li}, {et~al\mbox{.}}} \bibinfo{year}{2025}\natexlab{b}.
\newblock \showarticletitle{VITA-1.5: Towards GPT-4o Level Real-Time Vision and Speech Interaction}.
\newblock \bibinfo{journal}{\emph{arXiv preprint arXiv:2501.01957}} (\bibinfo{year}{2025}).
\newblock


\bibitem[Fu et~al\mbox{.}(2024d)]%
        {fu2024ocrbenchv2improvedbenchmark}
\bibfield{author}{\bibinfo{person}{Ling Fu}, \bibinfo{person}{Biao Yang}, \bibinfo{person}{Zhebin Kuang}, \bibinfo{person}{Jiajun Song}, \bibinfo{person}{Yuzhe Li}, \bibinfo{person}{Linghao Zhu}, \bibinfo{person}{Qidi Luo}, \bibinfo{person}{Xinyu Wang}, \bibinfo{person}{Hao Lu}, \bibinfo{person}{Mingxin Huang}, \bibinfo{person}{Zhang Li}, \bibinfo{person}{Guozhi Tang}, \bibinfo{person}{Bin Shan}, \bibinfo{person}{Chunhui Lin}, \bibinfo{person}{Qi Liu}, \bibinfo{person}{Binghong Wu}, \bibinfo{person}{Hao Feng}, \bibinfo{person}{Hao Liu}, \bibinfo{person}{Can Huang}, \bibinfo{person}{Jingqun Tang}, \bibinfo{person}{Wei Chen}, \bibinfo{person}{Lianwen Jin}, \bibinfo{person}{Yuliang Liu}, {and} \bibinfo{person}{Xiang Bai}.} \bibinfo{year}{2024}\natexlab{d}.
\newblock \bibinfo{title}{OCRBench v2: An Improved Benchmark for Evaluating Large Multimodal Models on Visual Text Localization and Reasoning}.
\newblock
\showeprint[arxiv]{2501.00321}~[cs.CV]
\urldef\tempurl%
\url{https://arxiv.org/abs/2501.00321}
\showURL{%
\tempurl}


\bibitem[Galvez et~al\mbox{.}(2021)]%
        {galvez2021peoplesspeechlargescalediverse}
\bibfield{author}{\bibinfo{person}{Daniel Galvez}, \bibinfo{person}{Greg Diamos}, \bibinfo{person}{Juan Ciro}, \bibinfo{person}{Juan~Felipe Cerón}, \bibinfo{person}{Keith Achorn}, \bibinfo{person}{Anjali Gopi}, \bibinfo{person}{David Kanter}, \bibinfo{person}{Maximilian Lam}, \bibinfo{person}{Mark Mazumder}, {and} \bibinfo{person}{Vijay~Janapa Reddi}.} \bibinfo{year}{2021}\natexlab{}.
\newblock \bibinfo{title}{The People's Speech: A Large-Scale Diverse English Speech Recognition Dataset for Commercial Usage}.
\newblock
\showeprint[arxiv]{2111.09344}~[cs.LG]
\urldef\tempurl%
\url{https://arxiv.org/abs/2111.09344}
\showURL{%
\tempurl}


\bibitem[Guan et~al\mbox{.}(2023)]%
        {guan2023hallusionbench}
\bibfield{author}{\bibinfo{person}{Tianrui Guan}, \bibinfo{person}{Fuxiao Liu}, \bibinfo{person}{Xiyang Wu}, \bibinfo{person}{Ruiqi Xian}, \bibinfo{person}{Zongxia Li}, \bibinfo{person}{Xiaoyu Liu}, \bibinfo{person}{Xijun Wang}, \bibinfo{person}{Lichang Chen}, \bibinfo{person}{Furong Huang}, \bibinfo{person}{Yaser Yacoob}, {et~al\mbox{.}}} \bibinfo{year}{2023}\natexlab{}.
\newblock \showarticletitle{HallusionBench: An Advanced Diagnostic Suite for Entangled Language Hallucination and Visual Illusion in Large Vision-Language Models}.
\newblock \bibinfo{journal}{\emph{arXiv preprint arXiv:2310.14566}} (\bibinfo{year}{2023}).
\newblock


\bibitem[Huh et~al\mbox{.}(2024)]%
        {pmlr-v235-huh24a}
\bibfield{author}{\bibinfo{person}{Minyoung Huh}, \bibinfo{person}{Brian Cheung}, \bibinfo{person}{Tongzhou Wang}, {and} \bibinfo{person}{Phillip Isola}.} \bibinfo{year}{2024}\natexlab{}.
\newblock \showarticletitle{Position: The Platonic Representation Hypothesis}. In \bibinfo{booktitle}{\emph{Proceedings of the 41st International Conference on Machine Learning}} \emph{(\bibinfo{series}{Proceedings of Machine Learning Research}, Vol.~\bibinfo{volume}{235})}, \bibfield{editor}{\bibinfo{person}{Ruslan Salakhutdinov}, \bibinfo{person}{Zico Kolter}, \bibinfo{person}{Katherine Heller}, \bibinfo{person}{Adrian Weller}, \bibinfo{person}{Nuria Oliver}, \bibinfo{person}{Jonathan Scarlett}, {and} \bibinfo{person}{Felix Berkenkamp}} (Eds.). \bibinfo{publisher}{PMLR}, \bibinfo{pages}{20617--20642}.
\newblock
\urldef\tempurl%
\url{https://proceedings.mlr.press/v235/huh24a.html}
\showURL{%
\tempurl}


\bibitem[Kaplan et~al\mbox{.}(2020)]%
        {kaplan2020scalinglawsneurallanguage}
\bibfield{author}{\bibinfo{person}{Jared Kaplan}, \bibinfo{person}{Sam McCandlish}, \bibinfo{person}{Tom Henighan}, \bibinfo{person}{Tom~B. Brown}, \bibinfo{person}{Benjamin Chess}, \bibinfo{person}{Rewon Child}, \bibinfo{person}{Scott Gray}, \bibinfo{person}{Alec Radford}, \bibinfo{person}{Jeffrey Wu}, {and} \bibinfo{person}{Dario Amodei}.} \bibinfo{year}{2020}\natexlab{}.
\newblock \bibinfo{title}{Scaling Laws for Neural Language Models}.
\newblock
\showeprint[arxiv]{2001.08361}~[cs.LG]
\urldef\tempurl%
\url{https://arxiv.org/abs/2001.08361}
\showURL{%
\tempurl}


\bibitem[Kornblith et~al\mbox{.}(2019)]%
        {kornblith2019similarity}
\bibfield{author}{\bibinfo{person}{Simon Kornblith}, \bibinfo{person}{Mohammad Norouzi}, \bibinfo{person}{Honglak Lee}, {and} \bibinfo{person}{Geoffrey Hinton}.} \bibinfo{year}{2019}\natexlab{}.
\newblock \showarticletitle{Similarity of neural network representations revisited}. In \bibinfo{booktitle}{\emph{International conference on machine learning}}. PMLR, \bibinfo{pages}{3519--3529}.
\newblock


\bibitem[LeCun et~al\mbox{.}(1995)]%
        {lecun1995convolutional}
\bibfield{author}{\bibinfo{person}{Yann LeCun}, \bibinfo{person}{Yoshua Bengio}, {et~al\mbox{.}}} \bibinfo{year}{1995}\natexlab{}.
\newblock \showarticletitle{Convolutional networks for images, speech, and time series}.
\newblock \bibinfo{journal}{\emph{The handbook of brain theory and neural networks}} \bibinfo{volume}{3361}, \bibinfo{number}{10} (\bibinfo{year}{1995}), \bibinfo{pages}{1995}.
\newblock


\bibitem[Li et~al\mbox{.}(2025)]%
        {li2025megrezomnitechnicalreport}
\bibfield{author}{\bibinfo{person}{Boxun Li}, \bibinfo{person}{Yadong Li}, \bibinfo{person}{Zhiyuan Li}, \bibinfo{person}{Congyi Liu}, \bibinfo{person}{Weilin Liu}, \bibinfo{person}{Guowei Niu}, \bibinfo{person}{Zheyue Tan}, \bibinfo{person}{Haiyang Xu}, \bibinfo{person}{Zhuyu Yao}, \bibinfo{person}{Tao Yuan}, \bibinfo{person}{Dong Zhou}, \bibinfo{person}{Yueqing Zhuang}, \bibinfo{person}{Shengen Yan}, \bibinfo{person}{Guohao Dai}, {and} \bibinfo{person}{Yu Wang}.} \bibinfo{year}{2025}\natexlab{}.
\newblock \bibinfo{title}{Megrez-Omni Technical Report}.
\newblock
\showeprint[arxiv]{2502.15803}~[cs.LG]
\urldef\tempurl%
\url{https://arxiv.org/abs/2502.15803}
\showURL{%
\tempurl}


\bibitem[Li et~al\mbox{.}(2024a)]%
        {li2024baichuan}
\bibfield{author}{\bibinfo{person}{Yadong Li}, \bibinfo{person}{Haoze Sun}, \bibinfo{person}{Mingan Lin}, \bibinfo{person}{Tianpeng Li}, \bibinfo{person}{Guosheng Dong}, \bibinfo{person}{Tao Zhang}, \bibinfo{person}{Bowen Ding}, \bibinfo{person}{Wei Song}, \bibinfo{person}{Zhenglin Cheng}, \bibinfo{person}{Yuqi Huo}, {et~al\mbox{.}}} \bibinfo{year}{2024}\natexlab{a}.
\newblock \showarticletitle{Baichuan-omni technical report}.
\newblock \bibinfo{journal}{\emph{arXiv preprint arXiv:2410.08565}} \bibinfo{volume}{2}, \bibinfo{number}{3} (\bibinfo{year}{2024}).
\newblock


\bibitem[Li et~al\mbox{.}(2024b)]%
        {li2024omnibenchfutureuniversalomnilanguage}
\bibfield{author}{\bibinfo{person}{Yizhi Li}, \bibinfo{person}{Ge Zhang}, \bibinfo{person}{Yinghao Ma}, \bibinfo{person}{Ruibin Yuan}, \bibinfo{person}{Kang Zhu}, \bibinfo{person}{Hangyu Guo}, \bibinfo{person}{Yiming Liang}, \bibinfo{person}{Jiaheng Liu}, \bibinfo{person}{Zekun Wang}, \bibinfo{person}{Jian Yang}, \bibinfo{person}{Siwei Wu}, \bibinfo{person}{Xingwei Qu}, \bibinfo{person}{Jinjie Shi}, \bibinfo{person}{Xinyue Zhang}, \bibinfo{person}{Zhenzhu Yang}, \bibinfo{person}{Xiangzhou Wang}, \bibinfo{person}{Zhaoxiang Zhang}, \bibinfo{person}{Zachary Liu}, \bibinfo{person}{Emmanouil Benetos}, \bibinfo{person}{Wenhao Huang}, {and} \bibinfo{person}{Chenghua Lin}.} \bibinfo{year}{2024}\natexlab{b}.
\newblock \bibinfo{title}{OmniBench: Towards The Future of Universal Omni-Language Models}.
\newblock
\showeprint[arxiv]{2409.15272}~[cs.CL]
\urldef\tempurl%
\url{https://arxiv.org/abs/2409.15272}
\showURL{%
\tempurl}


\bibitem[Liao et~al\mbox{.}(2024)]%
        {liao2024fishspeechleveraginglargelanguage}
\bibfield{author}{\bibinfo{person}{Shijia Liao}, \bibinfo{person}{Yuxuan Wang}, \bibinfo{person}{Tianyu Li}, \bibinfo{person}{Yifan Cheng}, \bibinfo{person}{Ruoyi Zhang}, \bibinfo{person}{Rongzhi Zhou}, {and} \bibinfo{person}{Yijin Xing}.} \bibinfo{year}{2024}\natexlab{}.
\newblock \bibinfo{title}{Fish-Speech: Leveraging Large Language Models for Advanced Multilingual Text-to-Speech Synthesis}.
\newblock
\showeprint[arxiv]{2411.01156}~[cs.SD]
\urldef\tempurl%
\url{https://arxiv.org/abs/2411.01156}
\showURL{%
\tempurl}


\bibitem[Lin et~al\mbox{.}(2024)]%
        {lin2024vilapretrainingvisuallanguage}
\bibfield{author}{\bibinfo{person}{Ji Lin}, \bibinfo{person}{Hongxu Yin}, \bibinfo{person}{Wei Ping}, \bibinfo{person}{Yao Lu}, \bibinfo{person}{Pavlo Molchanov}, \bibinfo{person}{Andrew Tao}, \bibinfo{person}{Huizi Mao}, \bibinfo{person}{Jan Kautz}, \bibinfo{person}{Mohammad Shoeybi}, {and} \bibinfo{person}{Song Han}.} \bibinfo{year}{2024}\natexlab{}.
\newblock \bibinfo{title}{VILA: On Pre-training for Visual Language Models}.
\newblock
\showeprint[arxiv]{2312.07533}~[cs.CV]
\urldef\tempurl%
\url{https://arxiv.org/abs/2312.07533}
\showURL{%
\tempurl}


\bibitem[Liu et~al\mbox{.}(2024)]%
        {liu2024llavanext}
\bibfield{author}{\bibinfo{person}{Haotian Liu}, \bibinfo{person}{Chunyuan Li}, \bibinfo{person}{Yuheng Li}, \bibinfo{person}{Bo Li}, \bibinfo{person}{Yuanhan Zhang}, \bibinfo{person}{Sheng Shen}, {and} \bibinfo{person}{Yong~Jae Lee}.} \bibinfo{year}{2024}\natexlab{}.
\newblock \bibinfo{title}{LLaVA-NeXT: Improved reasoning, OCR, and world knowledge}.
\newblock
\urldef\tempurl%
\url{https://llava-vl.github.io/blog/2024-01-30-llava-next/}
\showURL{%
\tempurl}


\bibitem[Liu et~al\mbox{.}(2023)]%
        {liu2023llava}
\bibfield{author}{\bibinfo{person}{Haotian Liu}, \bibinfo{person}{Chunyuan Li}, \bibinfo{person}{Qingyang Wu}, {and} \bibinfo{person}{Yong~Jae Lee}.} \bibinfo{year}{2023}\natexlab{}.
\newblock \bibinfo{title}{Visual Instruction Tuning}.
\newblock


\bibitem[Lu et~al\mbox{.}(2023)]%
        {lu2023mathvista}
\bibfield{author}{\bibinfo{person}{Pan Lu}, \bibinfo{person}{Hritik Bansal}, \bibinfo{person}{Tony Xia}, \bibinfo{person}{Jiacheng Liu}, \bibinfo{person}{Chunyuan Li}, \bibinfo{person}{Hannaneh Hajishirzi}, \bibinfo{person}{Hao Cheng}, \bibinfo{person}{Kai-Wei Chang}, \bibinfo{person}{Michel Galley}, {and} \bibinfo{person}{Jianfeng Gao}.} \bibinfo{year}{2023}\natexlab{}.
\newblock \showarticletitle{Mathvista: Evaluating mathematical reasoning of foundation models in visual contexts}.
\newblock \bibinfo{journal}{\emph{arXiv preprint arXiv:2310.02255}} (\bibinfo{year}{2023}).
\newblock


\bibitem[Lu et~al\mbox{.}(2024)]%
        {lu2024ovisstructuralembeddingalignment}
\bibfield{author}{\bibinfo{person}{Shiyin Lu}, \bibinfo{person}{Yang Li}, \bibinfo{person}{Qing-Guo Chen}, \bibinfo{person}{Zhao Xu}, \bibinfo{person}{Weihua Luo}, \bibinfo{person}{Kaifu Zhang}, {and} \bibinfo{person}{Han-Jia Ye}.} \bibinfo{year}{2024}\natexlab{}.
\newblock \bibinfo{title}{Ovis: Structural Embedding Alignment for Multimodal Large Language Model}.
\newblock
\showeprint[arxiv]{2405.20797}~[cs.CV]
\urldef\tempurl%
\url{https://arxiv.org/abs/2405.20797}
\showURL{%
\tempurl}


\bibitem[Luo et~al\mbox{.}(2025)]%
        {luo2025openomnilargelanguagemodels}
\bibfield{author}{\bibinfo{person}{Run Luo}, \bibinfo{person}{Ting-En Lin}, \bibinfo{person}{Haonan Zhang}, \bibinfo{person}{Yuchuan Wu}, \bibinfo{person}{Xiong Liu}, \bibinfo{person}{Min Yang}, \bibinfo{person}{Yongbin Li}, \bibinfo{person}{Longze Chen}, \bibinfo{person}{Jiaming Li}, \bibinfo{person}{Lei Zhang}, \bibinfo{person}{Yangyi Chen}, \bibinfo{person}{Hamid Alinejad-Rokny}, {and} \bibinfo{person}{Fei Huang}.} \bibinfo{year}{2025}\natexlab{}.
\newblock \bibinfo{title}{OpenOmni: Large Language Models Pivot Zero-shot Omnimodal Alignment across Language with Real-time Self-Aware Emotional Speech Synthesis}.
\newblock
\showeprint[arxiv]{2501.04561}~[cs.CL]
\urldef\tempurl%
\url{https://arxiv.org/abs/2501.04561}
\showURL{%
\tempurl}


\bibitem[Meta.(2024)]%
        {llama3.2}
\bibfield{author}{\bibinfo{person}{Meta.}} \bibinfo{year}{2024}\natexlab{}.
\newblock \showarticletitle{Llama3.2 technical report}.
\newblock  (\bibinfo{year}{2024}).
\newblock


\bibitem[Nachmani et~al\mbox{.}(2023)]%
        {nachmani2023spoken}
\bibfield{author}{\bibinfo{person}{Eliya Nachmani}, \bibinfo{person}{Alon Levkovitch}, \bibinfo{person}{Roy Hirsch}, \bibinfo{person}{Julian Salazar}, \bibinfo{person}{Chulayuth Asawaroengchai}, \bibinfo{person}{Soroosh Mariooryad}, \bibinfo{person}{Ehud Rivlin}, \bibinfo{person}{RJ Skerry-Ryan}, {and} \bibinfo{person}{Michelle~Tadmor Ramanovich}.} \bibinfo{year}{2023}\natexlab{}.
\newblock \showarticletitle{Spoken Question Answering and Speech Continuation Using Spectrogram-Powered LLM}.
\newblock \bibinfo{journal}{\emph{arXiv preprint arXiv:2305.15255}} (\bibinfo{year}{2023}).
\newblock


\bibitem[Nachmani et~al\mbox{.}(2024)]%
        {nachmani2024spokenquestionansweringspeech}
\bibfield{author}{\bibinfo{person}{Eliya Nachmani}, \bibinfo{person}{Alon Levkovitch}, \bibinfo{person}{Roy Hirsch}, \bibinfo{person}{Julian Salazar}, \bibinfo{person}{Chulayuth Asawaroengchai}, \bibinfo{person}{Soroosh Mariooryad}, \bibinfo{person}{Ehud Rivlin}, \bibinfo{person}{RJ Skerry-Ryan}, {and} \bibinfo{person}{Michelle~Tadmor Ramanovich}.} \bibinfo{year}{2024}\natexlab{}.
\newblock \bibinfo{title}{Spoken Question Answering and Speech Continuation Using Spectrogram-Powered LLM}.
\newblock
\showeprint[arxiv]{2305.15255}~[cs.CL]
\urldef\tempurl%
\url{https://arxiv.org/abs/2305.15255}
\showURL{%
\tempurl}


\bibitem[O'Neill et~al\mbox{.}(2021)]%
        {oneill2021spgispeech5000hourstranscribed}
\bibfield{author}{\bibinfo{person}{Patrick~K. O'Neill}, \bibinfo{person}{Vitaly Lavrukhin}, \bibinfo{person}{Somshubra Majumdar}, \bibinfo{person}{Vahid Noroozi}, \bibinfo{person}{Yuekai Zhang}, \bibinfo{person}{Oleksii Kuchaiev}, \bibinfo{person}{Jagadeesh Balam}, \bibinfo{person}{Yuliya Dovzhenko}, \bibinfo{person}{Keenan Freyberg}, \bibinfo{person}{Michael~D. Shulman}, \bibinfo{person}{Boris Ginsburg}, \bibinfo{person}{Shinji Watanabe}, {and} \bibinfo{person}{Georg Kucsko}.} \bibinfo{year}{2021}\natexlab{}.
\newblock \bibinfo{title}{SPGISpeech: 5,000 hours of transcribed financial audio for fully formatted end-to-end speech recognition}.
\newblock
\showeprint[arxiv]{2104.02014}~[cs.CL]
\urldef\tempurl%
\url{https://arxiv.org/abs/2104.02014}
\showURL{%
\tempurl}


\bibitem[Panayotov et~al\mbox{.}(2015a)]%
        {7178964}
\bibfield{author}{\bibinfo{person}{Vassil Panayotov}, \bibinfo{person}{Guoguo Chen}, \bibinfo{person}{Daniel Povey}, {and} \bibinfo{person}{Sanjeev Khudanpur}.} \bibinfo{year}{2015}\natexlab{a}.
\newblock \showarticletitle{Librispeech: An ASR corpus based on public domain audio books}. In \bibinfo{booktitle}{\emph{2015 IEEE International Conference on Acoustics, Speech and Signal Processing (ICASSP)}}. \bibinfo{pages}{5206--5210}.
\newblock
\href{https://doi.org/10.1109/ICASSP.2015.7178964}{doi:\nolinkurl{10.1109/ICASSP.2015.7178964}}


\bibitem[Panayotov et~al\mbox{.}(2015b)]%
        {panayotov2015librispeech}
\bibfield{author}{\bibinfo{person}{Vassil Panayotov}, \bibinfo{person}{Guoguo Chen}, \bibinfo{person}{Daniel Povey}, {and} \bibinfo{person}{Sanjeev Khudanpur}.} \bibinfo{year}{2015}\natexlab{b}.
\newblock \showarticletitle{Librispeech: an asr corpus based on public domain audio books}. In \bibinfo{booktitle}{\emph{2015 IEEE international conference on acoustics, speech and signal processing (ICASSP)}}. IEEE, \bibinfo{pages}{5206--5210}.
\newblock


\bibitem[Pratap et~al\mbox{.}(2020a)]%
        {pratap2020mls}
\bibfield{author}{\bibinfo{person}{Vineel Pratap}, \bibinfo{person}{Qiantong Xu}, \bibinfo{person}{Anuroop Sriram}, \bibinfo{person}{Gabriel Synnaeve}, {and} \bibinfo{person}{Ronan Collobert}.} \bibinfo{year}{2020}\natexlab{a}.
\newblock \showarticletitle{Mls: A large-scale multilingual dataset for speech research}.
\newblock \bibinfo{journal}{\emph{arXiv preprint arXiv:2012.03411}} (\bibinfo{year}{2020}).
\newblock


\bibitem[Pratap et~al\mbox{.}(2020b)]%
        {Pratap2020MLSAL}
\bibfield{author}{\bibinfo{person}{Vineel Pratap}, \bibinfo{person}{Qiantong Xu}, \bibinfo{person}{Anuroop Sriram}, \bibinfo{person}{Gabriel Synnaeve}, {and} \bibinfo{person}{Ronan Collobert}.} \bibinfo{year}{2020}\natexlab{b}.
\newblock \showarticletitle{MLS: A Large-Scale Multilingual Dataset for Speech Research}.
\newblock \bibinfo{journal}{\emph{ArXiv}}  \bibinfo{volume}{abs/2012.03411} (\bibinfo{year}{2020}).
\newblock


\bibitem[Radford et~al\mbox{.}(2022)]%
        {radford2022robustspeechrecognitionlargescale}
\bibfield{author}{\bibinfo{person}{Alec Radford}, \bibinfo{person}{Jong~Wook Kim}, \bibinfo{person}{Tao Xu}, \bibinfo{person}{Greg Brockman}, \bibinfo{person}{Christine McLeavey}, {and} \bibinfo{person}{Ilya Sutskever}.} \bibinfo{year}{2022}\natexlab{}.
\newblock \bibinfo{title}{Robust Speech Recognition via Large-Scale Weak Supervision}.
\newblock
\showeprint[arxiv]{2212.04356}~[eess.AS]
\urldef\tempurl%
\url{https://arxiv.org/abs/2212.04356}
\showURL{%
\tempurl}


\bibitem[Radford et~al\mbox{.}(2023)]%
        {radford2023robust}
\bibfield{author}{\bibinfo{person}{Alec Radford}, \bibinfo{person}{Jong~Wook Kim}, \bibinfo{person}{Tao Xu}, \bibinfo{person}{Greg Brockman}, \bibinfo{person}{Christine McLeavey}, {and} \bibinfo{person}{Ilya Sutskever}.} \bibinfo{year}{2023}\natexlab{}.
\newblock \showarticletitle{Robust speech recognition via large-scale weak supervision}. In \bibinfo{booktitle}{\emph{International conference on machine learning}}. PMLR, \bibinfo{pages}{28492--28518}.
\newblock


\bibitem[Raghu et~al\mbox{.}(2017)]%
        {raghu2017svcca}
\bibfield{author}{\bibinfo{person}{Maithra Raghu}, \bibinfo{person}{Justin Gilmer}, \bibinfo{person}{Jason Yosinski}, {and} \bibinfo{person}{Jascha Sohl-Dickstein}.} \bibinfo{year}{2017}\natexlab{}.
\newblock \showarticletitle{Svcca: Singular vector canonical correlation analysis for deep learning dynamics and interpretability}.
\newblock \bibinfo{journal}{\emph{Advances in neural information processing systems}}  \bibinfo{volume}{30} (\bibinfo{year}{2017}).
\newblock


\bibitem[Srinivasan et~al\mbox{.}(2021)]%
        {srinivasan2021wit}
\bibfield{author}{\bibinfo{person}{Krishna Srinivasan}, \bibinfo{person}{Karthik Raman}, \bibinfo{person}{Jiecao Chen}, \bibinfo{person}{Michael Bendersky}, {and} \bibinfo{person}{Marc Najork}.} \bibinfo{year}{2021}\natexlab{}.
\newblock \showarticletitle{Wit: Wikipedia-based image text dataset for multimodal multilingual machine learning}. In \bibinfo{booktitle}{\emph{Proceedings of the 44th international ACM SIGIR conference on research and development in information retrieval}}. \bibinfo{pages}{2443--2449}.
\newblock


\bibitem[Tong et~al\mbox{.}(2024)]%
        {tong2024cambrian1fullyopenvisioncentric}
\bibfield{author}{\bibinfo{person}{Shengbang Tong}, \bibinfo{person}{Ellis Brown}, \bibinfo{person}{Penghao Wu}, \bibinfo{person}{Sanghyun Woo}, \bibinfo{person}{Manoj Middepogu}, \bibinfo{person}{Sai~Charitha Akula}, \bibinfo{person}{Jihan Yang}, \bibinfo{person}{Shusheng Yang}, \bibinfo{person}{Adithya Iyer}, \bibinfo{person}{Xichen Pan}, \bibinfo{person}{Ziteng Wang}, \bibinfo{person}{Rob Fergus}, \bibinfo{person}{Yann LeCun}, {and} \bibinfo{person}{Saining Xie}.} \bibinfo{year}{2024}\natexlab{}.
\newblock \bibinfo{title}{Cambrian-1: A Fully Open, Vision-Centric Exploration of Multimodal LLMs}.
\newblock
\showeprint[arxiv]{2406.16860}~[cs.CV]
\urldef\tempurl%
\url{https://arxiv.org/abs/2406.16860}
\showURL{%
\tempurl}


\bibitem[Vaswani(2017)]%
        {vaswani2017attention}
\bibfield{author}{\bibinfo{person}{A Vaswani}.} \bibinfo{year}{2017}\natexlab{}.
\newblock \showarticletitle{Attention is all you need}.
\newblock \bibinfo{journal}{\emph{Advances in Neural Information Processing Systems}} (\bibinfo{year}{2017}).
\newblock


\bibitem[Wang et~al\mbox{.}(2020)]%
        {wang2020covost2massivelymultilingual}
\bibfield{author}{\bibinfo{person}{Changhan Wang}, \bibinfo{person}{Anne Wu}, {and} \bibinfo{person}{Juan Pino}.} \bibinfo{year}{2020}\natexlab{}.
\newblock \bibinfo{title}{CoVoST 2 and Massively Multilingual Speech-to-Text Translation}.
\newblock
\showeprint[arxiv]{2007.10310}~[cs.CL]
\urldef\tempurl%
\url{https://arxiv.org/abs/2007.10310}
\showURL{%
\tempurl}


\bibitem[Wang et~al\mbox{.}(2024)]%
        {wang2024qwen2}
\bibfield{author}{\bibinfo{person}{Peng Wang}, \bibinfo{person}{Shuai Bai}, \bibinfo{person}{Sinan Tan}, \bibinfo{person}{Shijie Wang}, \bibinfo{person}{Zhihao Fan}, \bibinfo{person}{Jinze Bai}, \bibinfo{person}{Keqin Chen}, \bibinfo{person}{Xuejing Liu}, \bibinfo{person}{Jialin Wang}, \bibinfo{person}{Wenbin Ge}, {et~al\mbox{.}}} \bibinfo{year}{2024}\natexlab{}.
\newblock \showarticletitle{Qwen2-vl: Enhancing vision-language model's perception of the world at any resolution}.
\newblock \bibinfo{journal}{\emph{arXiv preprint arXiv:2409.12191}} (\bibinfo{year}{2024}).
\newblock


\bibitem[Xie and Wu(2024a)]%
        {xie2024mini2}
\bibfield{author}{\bibinfo{person}{Zhifei Xie} {and} \bibinfo{person}{Changqiao Wu}.} \bibinfo{year}{2024}\natexlab{a}.
\newblock \showarticletitle{Mini-omni2: Towards open-source gpt-4o with vision, speech and duplex capabilities}.
\newblock \bibinfo{journal}{\emph{arXiv preprint arXiv:2410.11190}} (\bibinfo{year}{2024}).
\newblock


\bibitem[Xie and Wu(2024b)]%
        {xie2024mini}
\bibfield{author}{\bibinfo{person}{Zhifei Xie} {and} \bibinfo{person}{Changqiao Wu}.} \bibinfo{year}{2024}\natexlab{b}.
\newblock \showarticletitle{Mini-omni2: Towards open-source gpt-4o with vision, speech and duplex capabilities}.
\newblock \bibinfo{journal}{\emph{arXiv preprint arXiv:2410.11190}} (\bibinfo{year}{2024}).
\newblock


\bibitem[Xu et~al\mbox{.}(2025)]%
        {xu2025qwen25omnitechnicalreport}
\bibfield{author}{\bibinfo{person}{Jin Xu}, \bibinfo{person}{Zhifang Guo}, \bibinfo{person}{Jinzheng He}, \bibinfo{person}{Hangrui Hu}, \bibinfo{person}{Ting He}, \bibinfo{person}{Shuai Bai}, \bibinfo{person}{Keqin Chen}, \bibinfo{person}{Jialin Wang}, \bibinfo{person}{Yang Fan}, \bibinfo{person}{Kai Dang}, \bibinfo{person}{Bin Zhang}, \bibinfo{person}{Xiong Wang}, \bibinfo{person}{Yunfei Chu}, {and} \bibinfo{person}{Junyang Lin}.} \bibinfo{year}{2025}\natexlab{}.
\newblock \bibinfo{title}{Qwen2.5-Omni Technical Report}.
\newblock
\showeprint[arxiv]{2503.20215}~[cs.CL]
\urldef\tempurl%
\url{https://arxiv.org/abs/2503.20215}
\showURL{%
\tempurl}


\bibitem[Yang et~al\mbox{.}(2024b)]%
        {yang2024qwen2}
\bibfield{author}{\bibinfo{person}{An Yang}, \bibinfo{person}{Baosong Yang}, \bibinfo{person}{Beichen Zhang}, \bibinfo{person}{Binyuan Hui}, \bibinfo{person}{Bo Zheng}, \bibinfo{person}{Bowen Yu}, \bibinfo{person}{Chengyuan Li}, \bibinfo{person}{Dayiheng Liu}, \bibinfo{person}{Fei Huang}, \bibinfo{person}{Haoran Wei}, {et~al\mbox{.}}} \bibinfo{year}{2024}\natexlab{b}.
\newblock \showarticletitle{Qwen2. 5 Technical Report}.
\newblock \bibinfo{journal}{\emph{arXiv preprint arXiv:2412.15115}} (\bibinfo{year}{2024}).
\newblock


\bibitem[Yang et~al\mbox{.}(2024a)]%
        {yang2024air}
\bibfield{author}{\bibinfo{person}{Qian Yang}, \bibinfo{person}{Jin Xu}, \bibinfo{person}{Wenrui Liu}, \bibinfo{person}{Yunfei Chu}, \bibinfo{person}{Ziyue Jiang}, \bibinfo{person}{Xiaohuan Zhou}, \bibinfo{person}{Yichong Leng}, \bibinfo{person}{Yuanjun Lv}, \bibinfo{person}{Zhou Zhao}, \bibinfo{person}{Chang Zhou}, {et~al\mbox{.}}} \bibinfo{year}{2024}\natexlab{a}.
\newblock \showarticletitle{AIR-Bench: Benchmarking Large Audio-Language Models via Generative Comprehension}.
\newblock \bibinfo{journal}{\emph{arXiv preprint arXiv:2402.07729}} (\bibinfo{year}{2024}).
\newblock


\bibitem[Yao et~al\mbox{.}(2024)]%
        {yao2024minicpm}
\bibfield{author}{\bibinfo{person}{Yuan Yao}, \bibinfo{person}{Tianyu Yu}, \bibinfo{person}{Ao Zhang}, \bibinfo{person}{Chongyi Wang}, \bibinfo{person}{Junbo Cui}, \bibinfo{person}{Hongji Zhu}, \bibinfo{person}{Tianchi Cai}, \bibinfo{person}{Haoyu Li}, \bibinfo{person}{Weilin Zhao}, \bibinfo{person}{Zhihui He}, {et~al\mbox{.}}} \bibinfo{year}{2024}\natexlab{}.
\newblock \showarticletitle{MiniCPM-V: A GPT-4V Level MLLM on Your Phone}.
\newblock \bibinfo{journal}{\emph{arXiv preprint arXiv:2408.01800}} (\bibinfo{year}{2024}).
\newblock


\bibitem[Ye et~al\mbox{.}(2023)]%
        {ye2023gigast10000hourpseudospeech}
\bibfield{author}{\bibinfo{person}{Rong Ye}, \bibinfo{person}{Chengqi Zhao}, \bibinfo{person}{Tom Ko}, \bibinfo{person}{Chutong Meng}, \bibinfo{person}{Tao Wang}, \bibinfo{person}{Mingxuan Wang}, {and} \bibinfo{person}{Jun Cao}.} \bibinfo{year}{2023}\natexlab{}.
\newblock \bibinfo{title}{GigaST: A 10,000-hour Pseudo Speech Translation Corpus}.
\newblock
\showeprint[arxiv]{2204.03939}~[cs.CL]
\urldef\tempurl%
\url{https://arxiv.org/abs/2204.03939}
\showURL{%
\tempurl}


\bibitem[Yu et~al\mbox{.}(2024)]%
        {yu2024mmvetevaluatinglargemultimodal}
\bibfield{author}{\bibinfo{person}{Weihao Yu}, \bibinfo{person}{Zhengyuan Yang}, \bibinfo{person}{Linjie Li}, \bibinfo{person}{Jianfeng Wang}, \bibinfo{person}{Kevin Lin}, \bibinfo{person}{Zicheng Liu}, \bibinfo{person}{Xinchao Wang}, {and} \bibinfo{person}{Lijuan Wang}.} \bibinfo{year}{2024}\natexlab{}.
\newblock \bibinfo{title}{MM-Vet: Evaluating Large Multimodal Models for Integrated Capabilities}.
\newblock
\showeprint[arxiv]{2308.02490}~[cs.AI]
\urldef\tempurl%
\url{https://arxiv.org/abs/2308.02490}
\showURL{%
\tempurl}


\bibitem[Yue et~al\mbox{.}(2024)]%
        {yue2024mmmumassivemultidisciplinemultimodal}
\bibfield{author}{\bibinfo{person}{Xiang Yue}, \bibinfo{person}{Yuansheng Ni}, \bibinfo{person}{Kai Zhang}, \bibinfo{person}{Tianyu Zheng}, \bibinfo{person}{Ruoqi Liu}, \bibinfo{person}{Ge Zhang}, \bibinfo{person}{Samuel Stevens}, \bibinfo{person}{Dongfu Jiang}, \bibinfo{person}{Weiming Ren}, \bibinfo{person}{Yuxuan Sun}, \bibinfo{person}{Cong Wei}, \bibinfo{person}{Botao Yu}, \bibinfo{person}{Ruibin Yuan}, \bibinfo{person}{Renliang Sun}, \bibinfo{person}{Ming Yin}, \bibinfo{person}{Boyuan Zheng}, \bibinfo{person}{Zhenzhu Yang}, \bibinfo{person}{Yibo Liu}, \bibinfo{person}{Wenhao Huang}, \bibinfo{person}{Huan Sun}, \bibinfo{person}{Yu Su}, {and} \bibinfo{person}{Wenhu Chen}.} \bibinfo{year}{2024}\natexlab{}.
\newblock \bibinfo{title}{MMMU: A Massive Multi-discipline Multimodal Understanding and Reasoning Benchmark for Expert AGI}.
\newblock
\showeprint[arxiv]{2311.16502}~[cs.CL]
\urldef\tempurl%
\url{https://arxiv.org/abs/2311.16502}
\showURL{%
\tempurl}


\bibitem[Zeng et~al\mbox{.}(2024)]%
        {zeng2024scalingspeechtextpretrainingsynthetic}
\bibfield{author}{\bibinfo{person}{Aohan Zeng}, \bibinfo{person}{Zhengxiao Du}, \bibinfo{person}{Mingdao Liu}, \bibinfo{person}{Lei Zhang}, \bibinfo{person}{Shengmin Jiang}, \bibinfo{person}{Yuxiao Dong}, {and} \bibinfo{person}{Jie Tang}.} \bibinfo{year}{2024}\natexlab{}.
\newblock \bibinfo{title}{Scaling Speech-Text Pre-training with Synthetic Interleaved Data}.
\newblock
\showeprint[arxiv]{2411.17607}~[cs.CL]
\urldef\tempurl%
\url{https://arxiv.org/abs/2411.17607}
\showURL{%
\tempurl}


\bibitem[Zhang et~al\mbox{.}(2022)]%
        {zhang2022wenetspeech}
\bibfield{author}{\bibinfo{person}{Binbin Zhang}, \bibinfo{person}{Hang Lv}, \bibinfo{person}{Pengcheng Guo}, \bibinfo{person}{Qijie Shao}, \bibinfo{person}{Chao Yang}, \bibinfo{person}{Lei Xie}, \bibinfo{person}{Xin Xu}, \bibinfo{person}{Hui Bu}, \bibinfo{person}{Xiaoyu Chen}, \bibinfo{person}{Chenchen Zeng}, {et~al\mbox{.}}} \bibinfo{year}{2022}\natexlab{}.
\newblock \showarticletitle{Wenetspeech: A 10000+ hours multi-domain mandarin corpus for speech recognition}. In \bibinfo{booktitle}{\emph{ICASSP}}.
\newblock


\bibitem[Zhang et~al\mbox{.}(2023)]%
        {zhang2023speechgptempoweringlargelanguage}
\bibfield{author}{\bibinfo{person}{Dong Zhang}, \bibinfo{person}{Shimin Li}, \bibinfo{person}{Xin Zhang}, \bibinfo{person}{Jun Zhan}, \bibinfo{person}{Pengyu Wang}, \bibinfo{person}{Yaqian Zhou}, {and} \bibinfo{person}{Xipeng Qiu}.} \bibinfo{year}{2023}\natexlab{}.
\newblock \showarticletitle{SpeechGPT: Empowering Large Language Models with Intrinsic Cross-Modal Conversational Abilities}.
\newblock \bibinfo{journal}{\emph{arXiv preprint arXiv:2305.11000}} (\bibinfo{year}{2023}).
\newblock


\bibitem[Zhang et~al\mbox{.}(2024)]%
        {zhang-etal-2024-graph}
\bibfield{author}{\bibinfo{person}{Yingji Zhang}, \bibinfo{person}{Marco Valentino}, \bibinfo{person}{Danilo Carvalho}, \bibinfo{person}{Ian Pratt-Hartmann}, {and} \bibinfo{person}{Andre Freitas}.} \bibinfo{year}{2024}\natexlab{}.
\newblock \showarticletitle{Graph-Induced Syntactic-Semantic Spaces in Transformer-Based Variational {A}uto{E}ncoders}. In \bibinfo{booktitle}{\emph{Findings of the Association for Computational Linguistics: NAACL 2024}}, \bibfield{editor}{\bibinfo{person}{Kevin Duh}, \bibinfo{person}{Helena Gomez}, {and} \bibinfo{person}{Steven Bethard}} (Eds.). \bibinfo{publisher}{Association for Computational Linguistics}, \bibinfo{address}{Mexico City, Mexico}, \bibinfo{pages}{474--489}.
\newblock
\href{https://doi.org/10.18653/v1/2024.findings-naacl.32}{doi:\nolinkurl{10.18653/v1/2024.findings-naacl.32}}


\end{thebibliography}
\end{document}